%% file: main.tex
\documentclass[10pt,journal,compsoc,review]{IEEEtran}

\usepackage{multirow}
\usepackage{subcaption}
\usepackage{graphicx}
\usepackage{colortbl}
\usepackage{booktabs}
\usepackage{amsmath}
\usepackage{amssymb}
\usepackage{url}

\begin{document}

\title{AppGen: Mobility-aware App Usage Behavior Generation for Mobile Users}

\author{Zihan~Huang,
        Tong~Li,~\IEEEmembership{Member,~IEEE},
        Yong~Li,~\IEEEmembership{Member,~IEEE}
\IEEEcompsocitemizethanks
{\IEEEcompsocthanksitem Z. Huang, T. Li, and Y. Li are with the Beijing National Research Center for Information Science and Technology (BNRist), Department of Electronic Engineering, Tsinghua University, Beijing 100084, China. E-mail: tongli@mail.tsinghua.edu.cn
}
}

\markboth{IEEE TRANSACTIONS ON MOBILE COMPUTING}%
{Shell \MakeLowercase{\textit{et al.}}: Bare Advanced Demo of IEEEtran.cls for IEEE Computer Society Journals}
%

\IEEEtitleabstractindextext{%
\begin{abstract}
Mobile app usage behavior reveals human patterns and is crucial for stakeholders, but data collection is costly and raises privacy issues. Data synthesis can address this by generating artificial datasets that mirror real-world data. In this paper, we propose AppGen, an autoregressive generative model designed to generate app usage behavior based on users' mobility trajectories, improving dataset accessibility and quality. Specifically, AppGen employs a probabilistic diffusion model to simulate the stochastic nature of app usage behavior. By utilizing an autoregressive structure, AppGen effectively captures the intricate sequential relationships between different app usage events. Additionally, AppGen leverages latent encoding to extract semantic features from spatio-temporal points, guiding behavior generation. These key designs ensure the generated behaviors are contextually relevant and faithfully represent users' environments and past interactions. Experiments with two real-world datasets show that AppGen outperforms state-of-the-art baselines by over 12\% in critical metrics and accurately reflects real-world spatio-temporal patterns. We also test the generated datasets in applications, demonstrating their suitability for downstream tasks by maintaining algorithm accuracy and order.
\end{abstract}
\begin{IEEEkeywords}
Mobile app, app usage, data generation, diffusion model. 
\end{IEEEkeywords}
}

\maketitle
\IEEEdisplaynontitleabstractindextext
\IEEEpeerreviewmaketitle

\input{Introduction}

\input{Preliminaries}

\input{Methods}
\input{Methology}

\input{Evaluations}
\input{Application}
\input{Related}
\input{Conclusion}

\bibliographystyle{unsrt}
\bibliography{Ref}



\end{document}

%% file: Introduction.tex
\section{Introduction}

Mobile application (abbreviated as app) usage has surged as these apps empower individuals with information and multifaceted services, such as online shopping~\cite{gruning2023directing}, healthcare management~\cite{munzert2021tracking}, and social networking~\cite{jabeur2013mobile}. As of the third quarter of 2022, Android users could choose from 3.55 million apps on Google Play, while iOS users had access to approximately 1.6 million apps on the Apple App Store~\cite{statista}. Users can easily download, install, and access these apps to simplify their daily routines, expecting a fast and seamless experience~\cite{li2020extent}. 

Mobile app usage behavior reveals human app usage patterns and has significant implications for stakeholders, including smartphone manufacturers, network operators, app developers, and end consumers~\cite{li2022smartphone}. Analyzing these patterns allows smartphone manufacturers to optimize device performance by efficiently scheduling resources like memory and battery power, pre-loading frequently used apps, and releasing unnecessary background apps~\cite{shin2012understanding}. Network operators can utilize this data to dynamically adjust traffic offloading schemes and to plan and configure edge computing and storage infrastructure for optimizing latency for specific services~\cite{klas2017edge, paschos2018role}. App developers benefit from feedback on user needs and interests~\cite{li2021understanding}, enabling them to provide personalized services such as tailored recommendations and targeted advertisements, enhancing the quality of experience (QoE) and increasing profits.

Obtaining mobile app usage behavior data presents significant challenges due to high collection expenses and data privacy issues. Firstly, the collection process is complex, resource-intensive, and subject to strict local policies and regulations~\cite{zaeem2020effect}, which often limits access to a few researchers who have established collaborations with network operators or possess substantial resources for large-scale data collection. Also, gathering app usage data through crowdsourced methods requires considerable time, money, and technological infrastructure investments~\cite{li2020apps}.
Secondly, app usage data contains sensitive information, making the protection of user privacy paramount~\cite{zhao2019user, li2023you}. Sharing this data poses significant privacy risks and legal implications. To mitigate these risks, data providers often introduce perturbations, which degrade data quality and reduce its utility for downstream applications. 
As a result, researchers face substantial challenges in accessing and utilizing mobile app usage data. These obstacles hinder the progress and advancement of research in this area.

Data synthesis can be an effective solution for accessing high-quality data by generating artificial datasets replicating real-world data's statistical properties~\cite{kosta2012large, li2024mobile, yin2022practical}. This approach ensures the availability of large, diverse, and representative datasets while maintaining privacy and security. Motivated by these advantages and to overcome the challenges in collecting app usage behavior data, we aim to generate app usage behavior data for mobile users. Given that mobile app usage behavior is closely correlated with spatio-temporal context~\cite{yu2020semantic, li2021finding, zhao2016discovering}, we incorporate the mobility trajectories of mobile users to achieve more accurate and realistic generation results. Numerous studies have focused on synthetic human trajectory generation~\cite{feng2020learning, kulkarni2018generative, song2019generating}, providing a solid foundation for our research. Consequently, we focus on generating app usage behavior data based on users' mobility trajectories to enhance the accessibility and quality of the datasets. Nevertheless, achieving this goal entails addressing three main challenges.
\begin{itemize}
    \item \textbf{Inherent uncertainty in user behavior.} Human behavior, including app usage behavior, is inherently stochastic rather than deterministic. When users interact with apps, their behavior is influenced by various unobserved factors such as personal status or the external environment. These factors introduce uncertainty, making user behavior appear random. Unlike predictive tasks that seek to forecast specific outcomes, generating realistic app usage behavior data requires extensive modeling of this inherent randomness. Therefore, it is imperative to develop robust methodologies that can capture and represent the stochastic nature of user interactions with apps, ensuring the synthesized data accurately reflects the variability and complexity of real-world behaviors.
    \item \textbf{Intricate relationships between apps.} Multiple apps often contribute jointly to a user's current activity, making the sequential relationship between apps more complex than a simple pairwise interaction. For example, an individual might launch a fitness app before using a music app to accompany their workout. Similarly, they might open a note-taking app to jot down thoughts after reading online news and then use a social media app to share their notes with friends. These intricate sequences highlight the multifaceted nature of app usage patterns. Therefore, capturing and modeling these complex sequential interrelationships is essential to accurately generate mobile users' app usage behavior.
    \item \textbf{Complex spatio-temporal context.} Mobile app usage behavior is significantly affected by location and time of day. For instance, individuals typically use food delivery apps during mealtimes, listen to music while commuting, read the news at public transportation stations, and make online payments at malls. These examples illustrate how the spatio-temporal context shapes app usage patterns. Therefore, it is crucial to consider and model this context to accurately generate app usage behavior data, which involves understanding the typical locations and time-specific nature of app usage activities. By incorporating this detailed contextual information, we can create more realistic app usage behavior models that reflect the true complexity of users' daily interactions with their mobile devices.
\end{itemize}

To address the above challenges, we propose AppGen, an autoregressive generative model for mobility-aware app usage behavior generation for mobile Users. More specifically, AppGen employs the probabilistic diffusion model to model the stochastic nature of app usage behavior to solve the \textbf{first} challenge. 
By leveraging denoising operations, AppGen decomposes the original complex app usage data distribution into a multi-step Markov chain, where each step gradually transitions from a Gaussian noise distribution to a more structured form~\cite{ho2020denoising, rombach2022high}. This process allows AppGen to iteratively refine and generate realistic app usage patterns, effectively modeling the inherent stochastic nature present in real-world app usage data. Moreover, AppGen applies the autoregressive structure in the generation process to capture the intricate relationships across apps, solving the \textbf{second} challenge. 
The model generates each app in the sequence one at a time, using the previously generated app sequence as a condition to guide the generation of the current app. This iterative process enhances the model's ability to accurately reflect the dependencies across app usage events. Also, we design an attention-based historical behavioral feature encoder with masking and sliding mechanisms. This encoder effectively learns the representations of historically generated app usage behavior, facilitating the guidance for generating the following behavior. Furthermore, AppGen utilizes latent encoding to extract semantic features of spatio-temporal points, guiding behavior generation and addressing the \textbf{third} challenge. AppGen leverages an urban knowledge graph to depict the relationships between various urban elements, including base stations, regions, business areas, and points of interest (POIs). Through graph embedding, we extract the features of spatio-temporal points. These spatio-temporal features are then concatenated with generated app features, transforming the app usage data into spatio-temporal app usage points. Additionally, we design a conditional module to model the effect of historical spatio-temporal app usage behavior and current spatio-temporal features on current app usage. This ensures that the generated behaviors are contextually relevant and accurately reflect the user's environment and past spatio-temporal interactions.

\begin{figure*}[tb]
\vspace{-2mm}
\centering
\captionsetup{justification=centering}
    \begin{minipage}[t]{0.31\linewidth}
        \centering
        \includegraphics[width=1\linewidth]{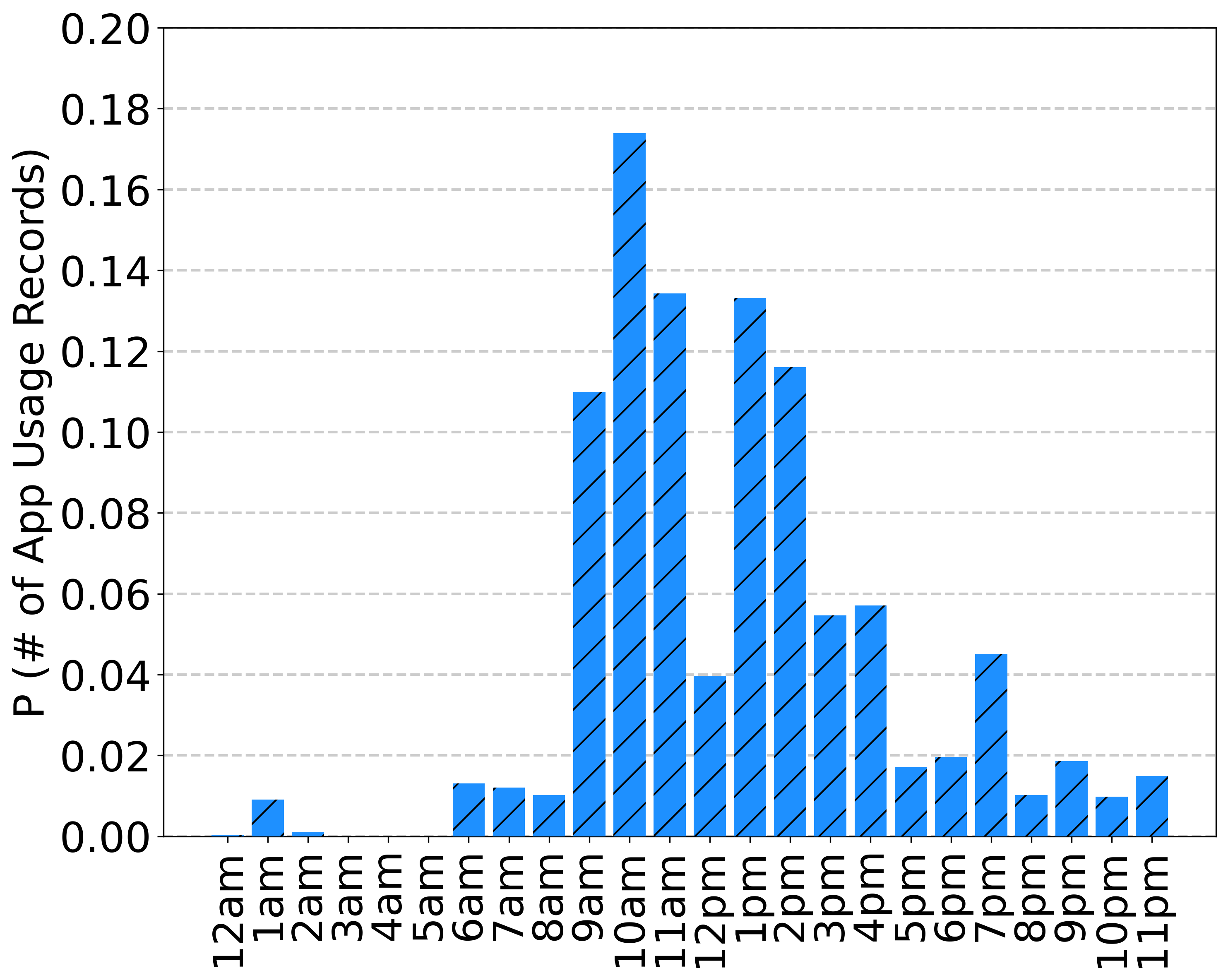}\\
        \subcaption{Finance}
        \label{fig:A1-1}
    \end{minipage}%
    \begin{minipage}[t]{0.31\linewidth}
        \centering
        \includegraphics[width=1\linewidth]{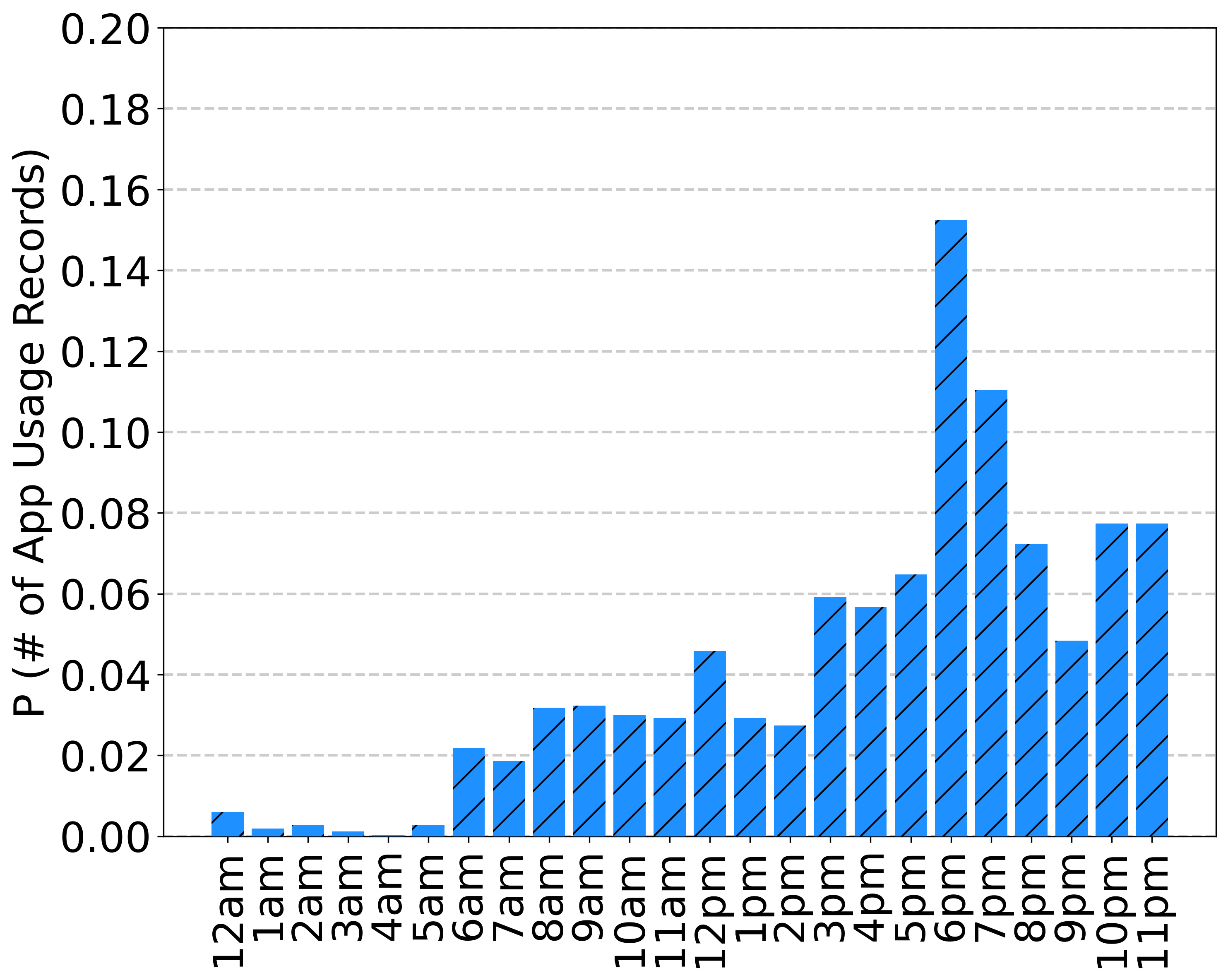}\\
        \subcaption{Games}
        \label{fig:A1-2}
    \end{minipage}%
    \begin{minipage}[t]{0.31\linewidth}
        \centering
        \includegraphics[width=1\linewidth]{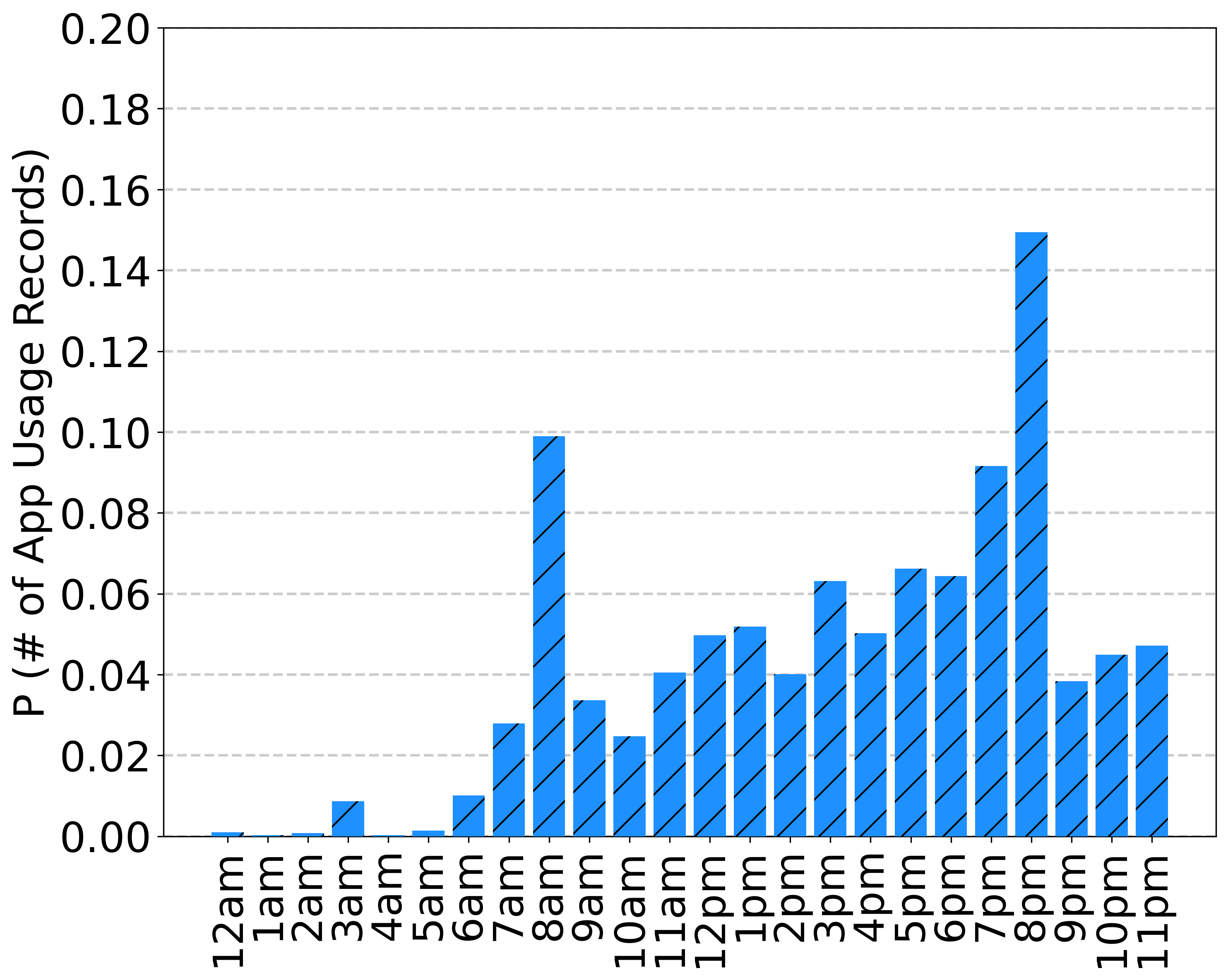}\\
        \subcaption{News}
        \label{fig:A1-3}
    \end{minipage}%
\vspace{-2mm}
\caption{App category usage records during a day based on Shanghai dataset.}
\vspace{-2mm} 
\label{fig:time_distribution}
\end{figure*}

Overall, the key contributions of our work can be summarized as follows:
\begin{itemize}  
    \item We investigate the problem of generating app usage behavior data based on users' mobility trajectories to enhance the accessibility and quality of datasets. To the best of our knowledge, we are the first to leverage conditional diffusion models for mobile app usage behavior generation. Our model incorporates the mobility trajectories of mobile users to achieve more accurate and realistic generation results, making it highly useful for context-aware downstream applications.
    \item We propose AppGen, an autoregressive generative model for mobility-aware app usage behavior generation for mobile Users. AppGen employs a probabilistic diffusion model to simulate the stochastic nature of app usage behavior. By utilizing an autoregressive structure, AppGen effectively captures the intricate sequential relationships between different app usage events during the generation process. Additionally, AppGen leverages latent encoding to extract semantic features from spatio-temporal points, guiding the behavior generation. This approach ensures that the generated behaviors are contextually relevant and faithfully represent the user's environment and past spatio-temporal interactions.
    \item We conduct extensive experiments using two real-world app usage behavior datasets to evaluate our model. The results demonstrate that AppGen significantly outperforms state-of-the-art baselines by over 12\% in terms of Jensen–Shannon divergence, Continuous Ranked Probability Score, and Spearman's Rank Correlation Coefficient. Moreover, the generated synthetic app usage behavior accurately follows the spatio-temporal patterns of real-world data. Additionally, we test the generated datasets in various applications, demonstrating that they meet users' requirements for downstream tasks by maintaining algorithm accuracy and order.
\end{itemize}

This paper is structured as follows. We begin by presenting preliminary studies in Section~\ref{sec:preliminary}. Next, we illustrate the overview of our problem and propose our autoregressive app usage behavior generative model based on diffusion models in Section~\ref{sec:method}. Following our methodology in Section~\ref{sec:methodology}, we present the evaluation and validation of our method in Section~\ref{sec:evaluate} as well as the application use cases in Section~\ref{sec:application}. Finally, after discussing important related work in Section~\ref{sec:related}, we conclude our paper in Section~\ref{sec:conclude}.

%% file: Preliminaries.tex
\section{Preliminaries}
\label{sec:preliminary}

\subsection{Contextual Impact on App Usage}

\begin{figure}[tb]
\vspace{-2mm}
\centering
\includegraphics[width=1.0\linewidth]{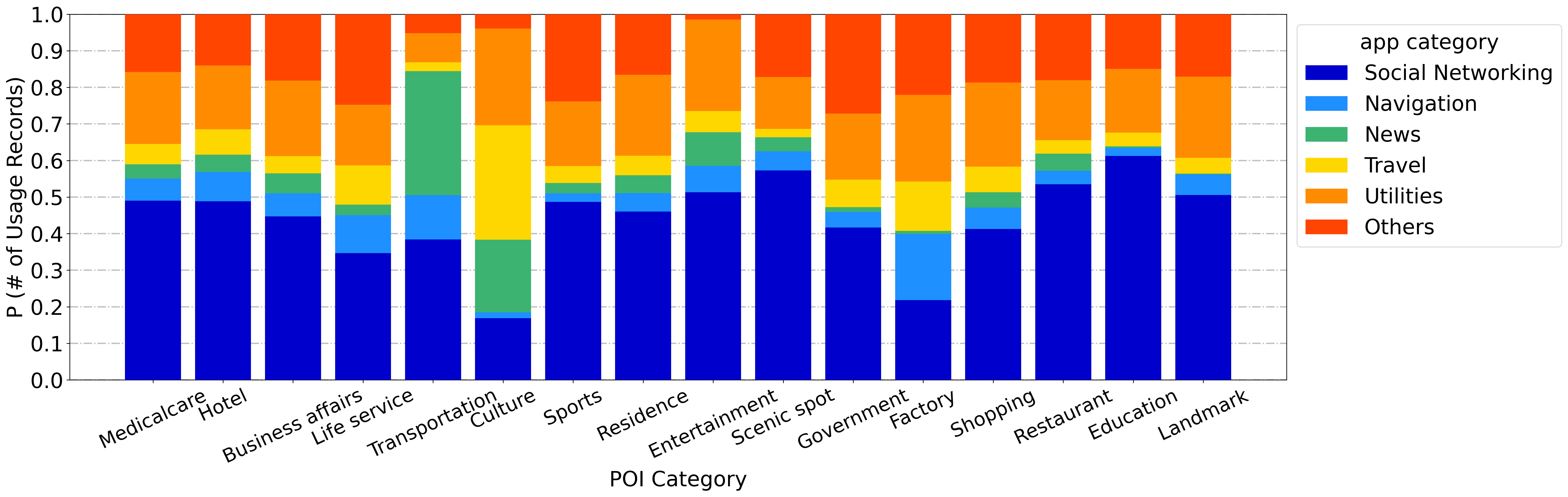}\\
\vspace{-2mm}
\caption{App category usage records distribution in terms of locations based on Shanghai dataset.}
\vspace{-2mm} 
\label{fig:loc_distribution}
\end{figure}

When utilizing apps, contextual information (e.g., time and location) significantly impacts users' behaviors.
Regarding temporal factors, user activities are often contingent on the time of day, and schedules are typically organized based on time, influencing app usage behavior. For example, during working hours, people are more likely to use business apps, whereas in the evening, they are more inclined to engage in leisure activities, making entertainment apps more popular. Some studies have confirmed this effect. Böhmer~\emph{et al.}~\cite{bohmer2011falling} found a higher probability of using news apps in the morning, finance apps around midday, and game apps at night.
Alternatively, spatial information and its different functional properties significantly influence app usage in varying ways. For example, Mehrotra~\emph{et al.}~\cite{mehrotra2017understanding} revealed that users primarily use apps for reading and listening to music while waiting at bus stops and train stations. Böhmer~\emph{et al.}~\cite{bohmer2011falling} also discovered that people were 2.78 times more likely to be using a browser and less likely to use reference, tool, or gaming apps in airports.

We also analyze a real-world mobile app usage dataset collected in Shanghai (details of the dataset are introduced in Section 3) and discover similar phenomena to those described above. \figurename~\ref{fig:time_distribution} shows the distribution of usage records for three app categories according to the hour of the day. We can observe that users generally start accessing apps between 7 and 8 a.m. From there, the frequency steadily rises to a peak and gradually decreases at night. Additionally, the utilization of different app categories varies significantly. Finance apps gain more popularity around midday or afternoon, whereas game apps are more frequently used in the evening, from 5 to 8 p.m. News apps exhibit two minor usage peaks, occurring around 8 a.m. and 8 p.m.
Furthermore, \figurename~\ref{fig:loc_distribution} illustrates the distribution of usage records for different app categories by location function over a day. We can see that travel apps are more prevalent in cultural settings, while apps used for navigation and news-checking are more frequently used in transportation-related areas. Communication apps dominate usage throughout most of the day. Therefore, we must incorporate spatio-temporal contextual information into the generation of mobile users' app usage behavior to accurately reflect the complexity of their daily interactions with mobile devices.

\subsection{Diffusion Models}

\begin{figure}[tb]
\vspace{-3mm}
\centering
\includegraphics[width=1.0\linewidth]{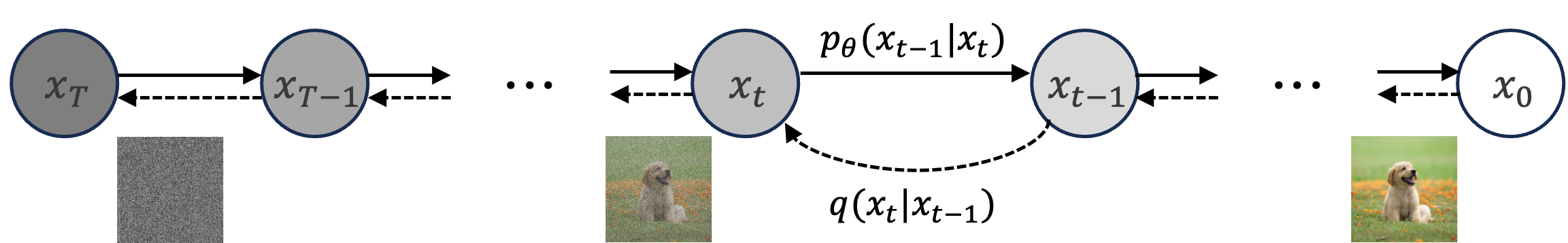}
\vspace{-3mm}
\caption{The graphical framework of the forward and reverse processes of the diffusion models.}
\label{fig:diffusion_model}
\vspace{-3mm}
\end{figure}

Diffusion models~\cite{sohl2015deep} are probabilistic generative models that generate samples consistent with the original data distribution. Specifically, diffusion models comprise two Markov chain processes--the forward and the reverse processes, as depicted in \figurename~\ref{fig:diffusion_model}.
Let $p_{\theta}({\bf x}_0)$ denote the synthetic data distribution that approximates the original data distribution $q({\bf x}_0)$. Let ${\bf x}_t$ for $t = 1,...,T$ represent a sequence of latent variables in the same sample space as ${\bf x}_0$, which is denoted as $\mathcal{X}$. The subsequent Markov chain defines the forward process:
\begin{equation}
    q({\bf x}_{1:T}|{\bf x}_0):=\prod_{t=1}^{T}q({\bf x}_{t}|{\bf x}_{t-1}),
\end{equation} 
where $q({\bf x}_{t}|{\bf x}_{t-1})=\mathcal{N}(\sqrt{1-\beta_{t}}{\bf x}_{t-1},\beta_t{\bf I})$ and $\beta_t$ is a small positive constant which represents Gaussian noise level. Equivalently, sampling ${\bf x}_t$ at an arbitrary timestep $t$ has a closed-form as:
\begin{equation}
    {\bf x}_t = \sqrt{\bar{\alpha}_t}{\bf x}_0+\sqrt{(1-\bar{\alpha}_t)}{\bf \epsilon},
\end{equation}
where we use the notation ${\alpha}_t:=(1-\beta_t)$, $\bar{\alpha}_t:=\prod_{s=1}^{t}{\alpha}_t$, and ${\bf \epsilon}\sim\mathcal{N}({\bf 0},{\bf I})$ is the Gaussian noise.
On the other hand, the reverse process denoises ${\bf x}_t$ to recover ${\bf x}_0$, which is defined by the Markov chain below:
\begin{equation}
    p_{\theta}({\bf x}_{0:T}):=p({\bf x}_T)\prod_{t=1}^{T}p_{\theta}({\bf x}_{t-1}|{\bf x}_t),
    \quad
    {\bf x}_T=\mathcal{N}({\bf 0},{\bf I}).
\end{equation}
Using a particular parametrization of $p_{\theta}({\bf x}_{t-1}|{\bf x}_t)$, Ho~\emph{et al.}~\cite{ho2020denoising} showed that the reverse process can be trained by optimizing the following loss function:
\begin{equation}
    \underset{\theta}{min}\mathcal{L}(\theta):=\underset{\theta}{min}\mathbb{E}_{{\bf x}_0\sim q({\bf x}_0),{\bf \epsilon}\sim \mathcal{N}({\bf 0},{\bf I}),t}||{\bf \epsilon}-{\bf \epsilon}_\theta({\bf x}_t,t)||_2^2.
    \label{diffloss}
\end{equation}

During sampling, we first sample noise points from a standard Gaussian distribution ${\bf z} \sim \mathcal{N}({\bf 0},{\bf I})$, and perform the reverse process starting from time step $T$ by using the learned noise prediction module ${\bf \epsilon}_\theta$ to estimate the added noise $\epsilon$ next. Then, we gradually eliminate the estimated noise from the noisy data to generate the synthetic data.

%% file: Methods.tex
\section{AppGen}
\label{sec:method}
In this section, we formalize the problem of mobility-aware app usage behavior generation, outline our AppGen approach, and present the detailed architecture of the whole model.

\subsection{Problem Statement}

\begin{figure}[tb]
\vspace{-3mm}
\centering
\includegraphics[width=1.0\linewidth]{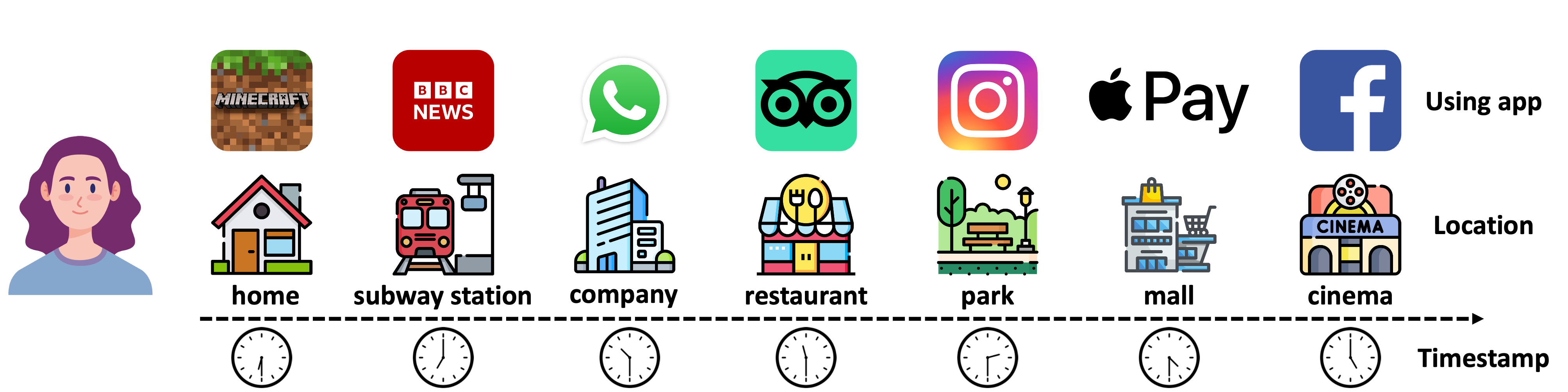}
\vspace{-3mm}
\caption{Spatio-temporal app usage behaviors of a user.}
\label{fig:ST_app}
\vspace{-3mm}
\end{figure}

Mobility-aware app usage behavior generation is to synthesize users' app usage sequences conditioned on spatio-temporal trajectories. 
Formally, let $\mathcal{A} = \{a_{1}, a_{2}, a_{3}, \dots, a_{N}\}$ denote the set of apps. Given a user $u$, his/her spatio-temporal app usage behaviors are defined as $X_u = \{A_u,S_u\}$, as shown in \figurename~\ref{fig:ST_app}. 
App usage sequence is formalized as $A_u = \{a_1,a_2,\dots,a_n\}$, where $n$ is the sequence length and $a_i$ represents the using app of the $i$-th usage. The mobility trajectory of user $u$ consists of a series of spatio-temporal points $S_u = \{s_1,s_2,\dots,s_n\}$, where each point is expressed as $s_i = (t_i, l_i)$, $t_i$ is the timestamp of the $i$-th visit, and $l_i$ indicates the location information. Given the mobility trajectory $S_u$, the main objective of app usage behavior generation is to generate the app usage sequence $\hat{A}_u$ such that the app usage probability distribution in the sequence matches the user's actual usage scenario $A_u$.

\subsection{Framework Overview}
\begin{figure*}[tb]
\vspace{-3mm}
\centering
\includegraphics[width=0.9\textwidth]{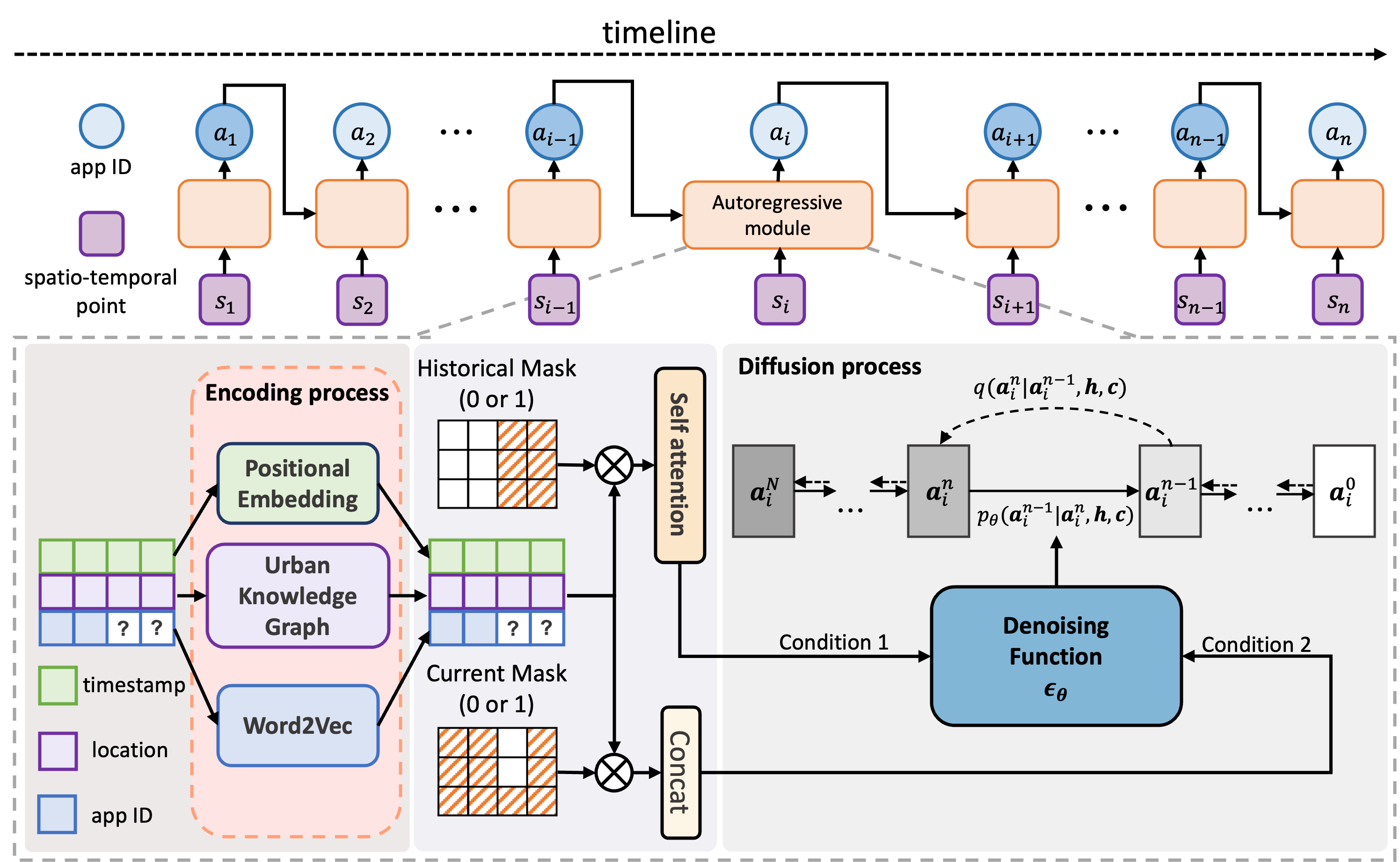}
\vspace{-3mm}
\caption{Framework overview of AppGen.}
\label{fig:framework}
\vspace{-3mm}
\end{figure*}

The framework of AppGen is shown in \figurename~\ref{fig:framework}. AppGen applies an autoregressive generation structure. More precisely, we generate each app in the sequence $A_u$ one by one. When generating the $i$-th app, the previously generated sequence, i.e., the historical app usage sequence of size $[1, i)$, will be part of the already known information to help guide the subsequent generation. In the autoregressive module, we apply the conditional diffusion model, where the previous sequence data $H$, including historical trajectory and app usage sequence, and current spatio-temporal information $C$, will be viewed as the supplemental information. Hence, the goal of diffusion models turns to estimate the conditional data distribution $q(A_u|H, C)$ with a model distribution $p_\theta(A_u|H, C)$, and we turn the reverse process into a conditional one:
\begin{equation}
p_{\theta}({\bf a}_{0:T}|H,C):=p({\bf a}_T)\prod_{t=1}^{T}p_{\theta}({\bf a}_{t-1}|{\bf a}_t,H,C).
\end{equation}

Specifically, AppGen is composed of three components:

\textbf{Latent representation encoding}. This process is responsible for mapping app IDs and spatio-temporal contextual information into latent spaces. These representations serve as conditions for the subsequent diffusion process.

\textbf{Historical behavioral feature extraction}. This involves capturing the historical behavioral characteristics of the user from their historical mobility trajectory and previously generated app usage sequences. These features will also serve as conditions to support the next app usage generation.

\textbf{Conditional app sequence generator}. This generates synthetic app usage sequence data conditioned on historical behavioral features and the current spatio-temporal context.

\subsection{Latent Representation Encoding}
Spatio-temporal app usage behavior involves different types of elements, including locations, timestamps, and apps, each with inherent relationships. In this module, we employ encoding techniques to map discrete element IDs within each type to a latent space that preserves the essential properties of the original space. This latent space efficiently captures these elements' relationships, facilitating the subsequent conditional diffusion processes.

\textbf{App Encoding.}
We project the discrete app IDs into a continuous latent space based on Word2Vec~\cite{mikolov2013efficient} algorithm to represent the inherent co-usage relationships of apps. Word2Vec was designed to represent words as high-dimension vectors that capture semantic information from surrounding words. 
In our case, we regard each app as a word and app usage sequences in the dataset as the corpus to learn one latent representation for each app in the whole app set, which can be expressed as $E^\mathcal{A} = \{{\bf a}_1, {\bf a}_2, \dots, {\bf a}_N\}$.

\textbf{Spatial Information Encoding.} 
We model the semantic meanings associated with various locations to fully develop how spatial context affects user behavior when using apps. 
We construct an urban knowledge graph to extract spatial factors. Specifically, a knowledge graph is a structure that views various contexts as entities, with knowledge being defined by the facts of various kinds of relations between entities~\cite{wang2021spatio}. Our urban knowledge graph is defined as $\mathcal{G}^{BS} = (\mathcal{E}^{BS}, \mathcal{R}^{BS}, \mathcal{F}^{BS})$, where $\mathcal{E}^{BS}$, $\mathcal{R}^{BS}$, and $\mathcal{F}^{BS}$ represent the sets of entities, relations, and facts respectively. Urban components consist of base stations, regions, business areas, and POIs are modeled as entities, and their spatial correlations are modeled as relations.  

\begin{figure}[tb]
\vspace{-2mm}
\centering
\includegraphics[width=0.92\linewidth]{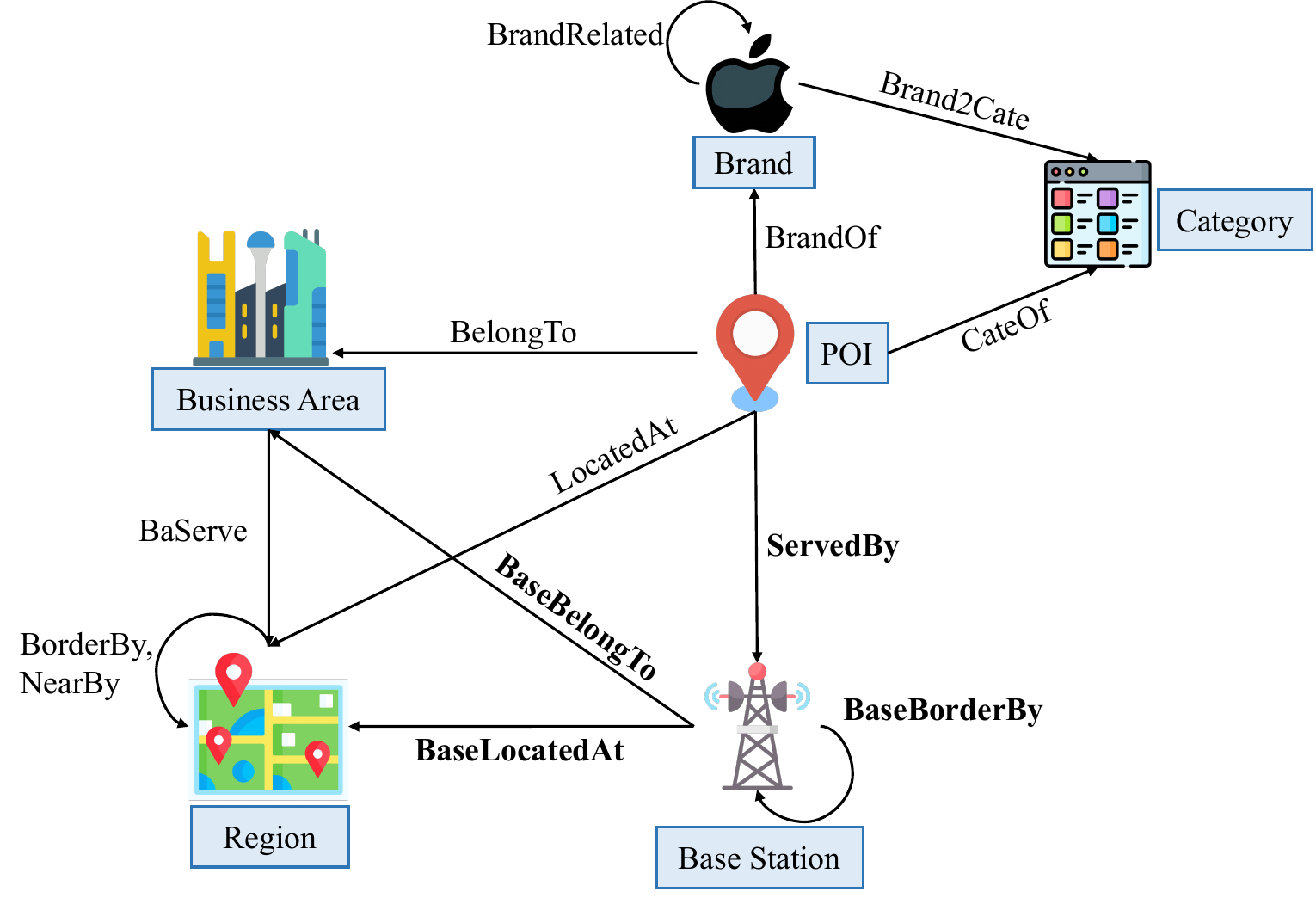}
\vspace{-2mm}
\caption{Constructed urban knowledge graph to acquire spatial contextual representations.}
\label{fig:KG}
\vspace{-2mm}
\end{figure}

The app usage datasets, in our case, are collected through mobile networks; thus, we treat base stations as locations. As shown in \figurename~\ref{fig:KG}, a base station is connected to other entities through four different kinds of relations: $(1)$ a base station is located at region--\textbf{BaseLocateAt}, $(2)$ a base station belongs to a business area--\textbf{BaseBelongTo}, $(3)$ a base station serves a POI--\textbf{ServedBy}, $(4)$ a base station borders with another base station--\textbf{BaseBorderBy}. 
Then, we apply TuckER~\cite{balavzevic2019tucker}, a state-of-art tensor factorization technique, for knowledge graph embedding, which maps each entity to a low dimensional vector while maintaining their semantic meanings, aiming to extract the spatial features modeled in the knowledge graph extensively. The TuckER model employs the cross-entropy loss function to optimize the embedding vectors of graph entities by measuring the plausibility of facts in the knowledge graph. 
At last, we acquire the well-learned spatial contextual embedding vectors $E^{BS} = \{{\bf l}_1, {\bf l}_2,\dots\}$, which can be used in further app usage generation.

\textbf{Temporal Information Encoding.} 
We apply timestamps to extract the influence of temporal context on app usage behavior. More precisely, we divide a day into forty-eight half-an-hour time bins and then encode these time bins into 128-dimensional temporal embedding following previous studies~\cite{vaswani2017attention, zuo2020transformer}: 
\begin{equation}
\begin{split}
    {\bf t}_i = (& sin(i/\tau^{0/64}),\dots,sin(i/\tau^{63/64}), \\
    & cos(i/\tau^{0/64}),\dots,cos(i/\tau^{63/64})),
\end{split}
\end{equation}
where $\tau=10000$, and we obtain the temporal representations denoted as $E^T = \{{\bf t}_1, {\bf t}_2,\dots,{\bf t}_{48}\}$.

\subsection{Historical Behavioral Feature Extraction}
Since previous app usage behavior significantly influences a user's current behavior, extracting historical behavioral features for app usage behavior generation is critical. To achieve this, we first utilize masking and sliding-window techniques to extract effective historical behavior information. We then apply an attention mechanism to capture the underlying behavioral characteristics.

\textbf{Spatio-temporal Context Extracting.} 
Since only past behavior influences the user's current decision-making when using apps and their future trajectory is unknowable, it is necessary and reasonable to mask the future trajectory during the generation process. Considering generating the $i$-th app in the app usage sequence, the user's past trajectory should consist of the previous $(i-1)$ spatio-temporal points. For further explanation, we denote a historical mask ${\bf M} = \{ m_{1:n}\} \in \mathbb{R}^{n}$, where $m_k = 1$ if $k < i$, otherwise $m_k = 0$. Then, the user's past trajectory can be expressed as:
\begin{equation}
    S_{u,past} = S_u \odot {\bf M} = \{s_1,s_2,\dots,s_{i-1}\},
\end{equation}
where $s_k = (t_k,l_k)$.

Nevertheless, not all historical behaviors are reliable information, and those too distant from the present may contribute noise to the analysis and make it more challenging to extract historical behavioral features related to the user's current app usage behavior. To address this issue, we employ a sliding window of fixed size $k$ to extract the previous app usage behavior of length $k$ for the current $i$-th app to be generated. Formally, the effective user's historical trajectory will be expressed as:
\begin{equation}
    S_{u,past} = \{s_{i-k},s_{i-k+1},\dots,s_{i-1}\},
\end{equation}

\textbf{Self Attention-based Encoding.} 
Due to varying intentions, users exhibit diverse behavioral trajectories, which have differing effects on the current app generation. We need to extract a historical behavioral trait from them to direct the generation task. In this case, we apply an attention mechanism to fuse and capture the user's historical app usage behavioral characteristics. 
Given the past trajectory $S_{u,past} = \{s_{i-k},s_{i-k+1},\dots,s_{i-1}\}$ and the historically generated app usage sequence $\{a_{i-k}, a_{i-k+1}, \dots, a_{i-1}\}$, we first find the corresponding temporal representations $\{{\bf t}_{i-k}, {\bf t}_{i-k+1}, \dots, {\bf t}_{i-1}\}$ from $E^{T}$, spatial representations $\{{\bf l}_{i-k}, {\bf l}_{i-k+1}, \dots, {\bf l}_{i-1}\}$ from $E^{BS}$ and app embeddings $\{{\bf a}_{i-k}, {\bf a}_{i-k+1}, \dots, {\bf a}_{i-1}\}$ from $E^{\mathcal{A}}$, respectively. Next, we concatenate the temporal, spatial, and app embedding vectors of each app usage sequence point, which is shown below: 
\begin{equation}
    {\bf h}_j = [{\bf t}_j, {\bf l}_j, {\bf a}_j], \quad j=i-k,i-k+1,\dots,i-1,
\end{equation}
and we obtain the behavioral feature representation of each sequence point ${\bf H}_u = \{{\bf h}_{i-k}, \dots, {\bf h}_{i-1}\}$.

Then, we adopt the self-attention mechanism to extract the historical behavioral feature, which is calculated as:
\begin{equation}
    \gamma_p = \frac{S({\bf W}_q{\bf h}_{i-1},{\bf W}_k{\bf h}_j)}{\sum_{i-k\leq j\leq i-1}S({\bf W}_q{\bf h}_{i-1},{\bf W}_k{\bf h}_j)},
\end{equation}

\begin{equation}
    \tilde{\bf h}_i = \sum_{i-k\leq p\leq i-1}\gamma_p{\bf W}_v{\bf h}_p,
\end{equation}
where the function $S(\cdot,\cdot)$ is $Scaled\,Dot$-$Product\,Attention$. ${\bf W}_q$, ${\bf W}_k$ and ${\bf W}_v$ are all trainable matrices. At this stage, we get the historical behavioral representation $\tilde{\bf h}_i$ for generating the $i$-th app in the app sequence.

\begin{figure}[tb]
\vspace{-2mm}
\centering
\includegraphics[width=0.92\linewidth]{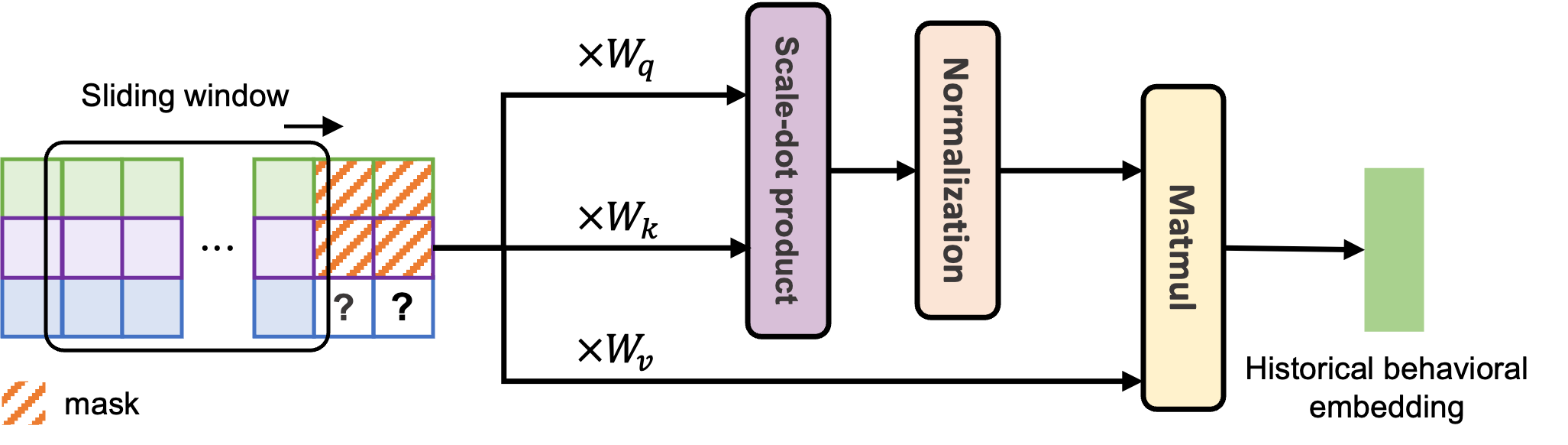}
\vspace{-2mm}
\caption{Extracting historical behavioral features through self-attention mechanism.}
\label{fig:history feature}
\vspace{-2mm}
\end{figure}

\subsection{Conditional App Sequence Generator}
\begin{figure}[tb]
\vspace{-2mm}
\centering
\includegraphics[width=0.92\linewidth]{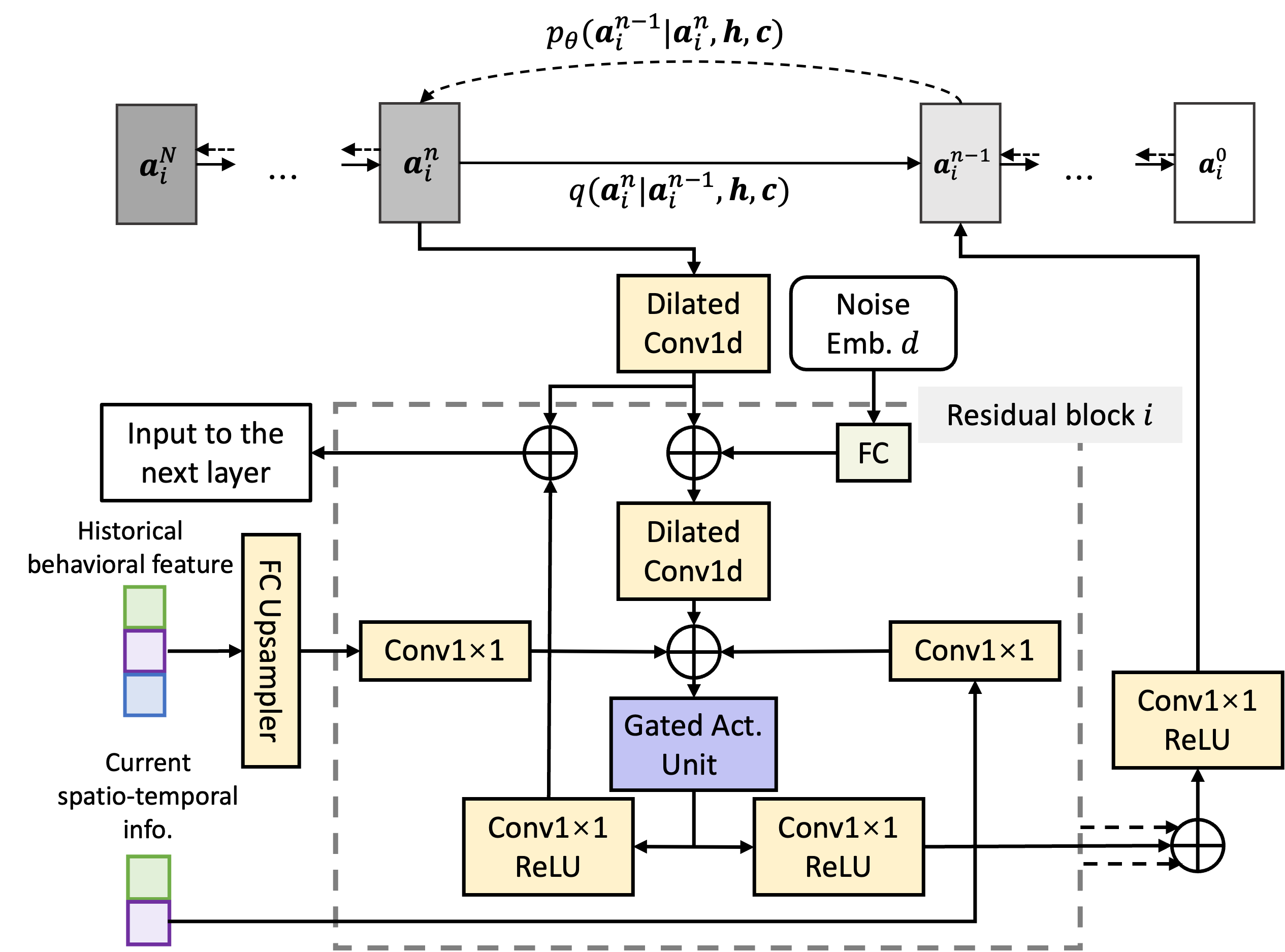}
\caption{Diffusion process schematic.}
\label{fig:diffusion process}
\vspace{-2mm}
\end{figure}

Since the goal is to generate the synthetic data of the app sequence, we utilize the conditional diffusion models to conduct the task. As the diffusion process is conducted in the latent space, we generate the corresponding app's representation, expecting it to be as close as feasible to its inherent embedding vectors. 
For the denoising network of diffusion models, we adopt the architecture in DiffWave~\cite{kong2020diffwave}, which is composed of multiple residual layers, as the neural backbone base of our model. 
The denoising network is conditioned with two auxiliary information--historical behavioral features and current spatio-temporal context. For the former, we pass the historical behavioral representation $\tilde{\bf h}_i$ through the linear layers to up/down-sample them to the appropriate size before incorporating them into the diffusion model. For the latter one, we first concatenate and fuse the temporal and spatial representations of the current trajectory point and then add them to the diffusion process after up/down sampling.
\figurename~\ref{fig:diffusion process} shows the schematics of a single residual block $i=\{1,2,\dots,8\}$ as well as the ﬁnal output from the sum of all the eight skip-connections. 
Every single residual block contains 1-dim dilated ConvNets and multiple 1D convolutional layers. For the noise index $t \in \{1,2,\dots, T\}$, we encode it by applying the Transformer's Fourier positional embeddings~\cite{vaswani2017attention} into vectors ${\bf d} \in \mathbb{R}^{32}$. 

Through the denoising function $\epsilon_{\theta}$, the noise added to its noisy input $\hat{\bf a}_i^t$ is estimated. We progressively reduce the noise in the noisy data using the acquired noise until we obtain the generated synthetic app embedding vector $\hat{\bf a}_i$. Ultimately, we compare the synthetic app embedding $\hat{\bf a}_i$ with the inherent app embedding of every app through the cosine similarity function, and the one with the highest similarity is the $i$-th app.

\subsection{Loss Functions}
Since the $\epsilon_\theta$ network is conditioned on the historical behavioral traits and current spatio-temporal context, we have that the conditional variant of the objective (\ref{diffloss}) for time step $i$ and noise index $t$:
\begin{equation}
    \mathcal{L}_{diff} = \mathbb{E}_{{\bf a}_i^0,\epsilon}||\epsilon-\epsilon_\theta({\bf a}_i^t,{\bf h}_i, {\bf c}_i, t)||_2^2,
    \quad
    {\bf a}_i^t = \sqrt{\bar{\alpha}_t}{\bf a}_i^0+\sqrt{1-\bar{\alpha}_t}{\bf \epsilon}.
\end{equation}
Besides, we also propose an embedding loss to constrain the difference between generated app embedding and real embedding vectors, which is formulated as:
\begin{equation}
    \mathcal{L}_{emb} = ||{\bf a}_i - \hat{\bf a}_i||_2^2,
\end{equation}
where ${\bf a}_i$ is the inherent embedding of app $a_i$ and $\hat{\bf a}_i$ is the synthetic embedding.

Furthermore, we also find that applying these losses uniformly over all diffusion steps is inappropriate. When $t$ gets close to the total time step $T$, there will be enough noise to distort the app embedding input into a noised version that has nearly lost all its valid information. To tackle this, we introduce a dynamic loss decay weight $\lambda_t = 1 - \alpha\frac{t}{T}$, which reduces linearly as diffusion step $t$ increases. The overall training loss is expressed as follows:
\begin{equation}
    \mathcal{L} = \mathcal{L}_{diff} + \lambda_t \mathcal{L}_{emb}.
\end{equation}

\begin{table*}[tb]
\centering
\caption{Dataset Statistics.}
\vspace{-2mm}
\footnotesize
{
    \begin{tabular}{c|cccccc}
    \toprule
    \multirow{2}{*}{\textbf{Cities}} & \multirow{2}{*}{\textbf{Collection Date}} & \multirow{2}{*}{\textbf{\# Users}} & \multirow{2}{*}{\textbf{\# Apps}}  & \multirow{2}{*}{\textbf{\# Records}} & \textbf{\# App} & \textbf{Avg. App Seq.} \\
    & & & & & \textbf{Categories} & \textbf{Length per day} \\
    \midrule 
    \textbf{Shanghai} & 19 April 2016 - 25 April 2016 & 11158 & 2000 &  28,797,454 & 20 & 492.6 \\
    \textbf{Nanchang} & 19 May 2022 - 25 May 2022 & 2000 & 1160 & 19,929,538 & 22 & 1057.2 \\
    \bottomrule
    \end{tabular}
    \label{Tab1}
}%
\end{table*}

\subsection{App Usage Generating}
To generate the synthetic app usage sequence $\hat{A}_u$ with trained AppGen, we take the $i$-th using app generation as an example. We first encode the apps and spatio-temporal context to obtain their latent representations. Next, we capture the historical app usage behavioral features via the self-attention mechanism. Then, we can generate the $i$-th app starting from Gaussian noise $\hat{\bf a}_i^T\sim \mathcal{N}({\bf 0},{\bf I})$ conditioned on previous sequence data $\tilde{\bf h}_i$ and current spatio-temporal information ${\bf c}_i$. Specifically, the reconstruction of $\hat{a}^0_i$ can be formulated as follows:
\begin{equation}
    \hat{\bf a}_i^{t-1}=\frac{1}{\sqrt{\alpha_t}}(\hat{\bf a}_i^t-\frac{\beta_t}{\sqrt{1-\bar{\alpha}_t}}\epsilon_\theta(\hat{\bf a}_i^t, \tilde{\bf h}_i, {\bf c}_i)) + \sqrt{\beta_k}{\bf z},
\end{equation} 
where ${\bf z}$ is a stochastic variable sampled from a standard Gaussian distribution, and $\epsilon_\theta$ is the trained reverse denoising network. By applying the above procedure to generate apps one at a time, we can finally obtain the entire app usage sequence.

%% file: Methology.tex
\section{Evaluation Methodology}
\label{sec:methodology}
In this section, we first introduce the dataset used to test our model. Next, we illustrate several metrics applied in our evaluation. Finally, we elaborate various baselines for comparison.

\subsection{Dataset Overview}
In the following experiments, we utilize two real-world datasets to measure the performance of our model. An Internet service provider in China collected one during a single week in April 2016. Shanghai, one of the biggest cities in the world, and its whole metropolitan region are all included in the dataset. 
The other dataset was collected by China Mobile, containing 19,929,538 records covering the entire metropolitan area of Nanchang, a capital city in China. The dataset was collected from the 19th of May 2022 to the 25th of May 2022. 
Both datasets' records include an anonymous user ID, an app type ID, an app ID, the timestamp.  
We split both datasets into the training set, validation set, and testing set in a ratio of 7:1:2 and ensured no overlap between them.
Details about the datasets are displayed in Table~\ref{Tab1}.

\subsection{Evaluation Metrics}
We assess the fidelity of AppGen with the baseline models via six different quantitative metrics, outlined below, which cover a wide range of aspects of interest:

\textbf{Root Mean Square Error (RMSE)}. This metric measures the average magnitude of the errors between generated values and actual values, which is defined as:
\begin{equation}
RMSE = \sqrt{\frac{1}{n}\sum_{i=1}^{n}(x_i - \hat{x}_i)^2}
\end{equation}
where $x_i$ is the actual value of the $i$-th sample and $\hat{x}_i$ is the  synthetic value.

\textbf{Mean Absolute Error (MAE)}. This metric is used to evaluate the average discrepancy between generated and real data; the formula is as follows:
\begin{equation}
RMSE = \frac{1}{n}\sum_{i=1}^{n}|x_i - \hat{x}_i|
\end{equation}
where $x_i$ and $\hat{x}_i$ are the original and generated values of the $i$-th sample respectively.

\textbf{Continuous Ranked Probability Score (CRPS)}. This metric is a scoring function that measures the compatibility between a cumulative distribution function $F$ with an observation $x$, which can be computed as:
\begin{equation}
CRPS(F,x) = \int (F(z) - {\mathbb{I}}\{x \leq z\})^2 dz
\end{equation}
where ${\mathbb{I}}\{x \leq z\}$ is the indicator function which equals to one if $x \leq z$ and zero otherwise. Thus, CRPS reaches its minimum when the generated data distribution $F$ and the ground-truth data distribution are equal.

\textbf{Jensen-Shannon divergence (JSD)~\cite{menendez1997jensen}}. This metric is a method of measuring the similarity between two probability distributions, and it can be calculated as:
\begin{equation}
JSD = \frac{1}{2}{\bf KL}(P\:||\:M)+\frac{1}{2}{\bf KL}(Q\:||\:M)
\end{equation}
where $P$ and $Q$ are distributions of generated and real data, $M=\frac{1}{2}(P+Q)$ is the mixture distribution, and ${\bf KL}(\cdot)$ is the Kullback-Leibler divergence.

\textbf{Marginal by total variation (M-TV)}. This metric measures how closely the generated data's distribution resembles the real data, which can be expressed as follows:
\begin{equation}
M\text{-}TV = \frac{1}{2}(|P - Q|)
\end{equation} 
where $P$ and $Q$ are marginal distributions for generated and real data, respectively.

\textbf{Spearman's Rank Correlation Coefficient(Spearmanr)}. This metric is a statistical measure used to quantify the association strength between two ranked variables, and it is formulated as follows: 
\begin{equation}
\rho = \frac{cov(R(X),R(Y))}{\sigma_{R(X)}\sigma_{R(Y)}}
\end{equation} 
where $X$ and $Y$ are two variables, $R(X)$ and $R(Y)$ are their ranks.

As for RMSE, MAE, JSD, CRPS, and M-TV, lower values indicate a better match between generated and real-world data. In contrast, a larger value is preferred for Spearmanr.

\begin{table*}[tb]
\centering
\caption{App usage generation results compared with baselines on the Shanghai dataset.}
\footnotesize
\resizebox{1.0\linewidth}{!}
{
    \begin{tabular}{c|ccccccccccccc}
    \toprule
    \multirow{2}{*}{\textbf{Methods}} & \multicolumn{6}{c}{{\textbf{App Popularity Distribution}}} & \multicolumn{6}{c}{\textbf{App Cate. Popularity Distribution}} \\
    \cmidrule(r){2-7}
    \cmidrule(r){8-13}
     & {RMSE} & {MAE} & {JSD} & {CRPS} & {M-TV} &{Spearmanr} & {RMSE} & {MAE} & {JSD} & {CRPS} & {M-TV} &{Spearmanr} \\
    \midrule
    Time-GAN&{4.9023}&{4.7846}&{0.0012}&{1.6994}&{2.3923}&{0.5233}&{4.8230}&{4.6178}&{0.0285}&{1.7112}&{2.3089}&{0.5581}\\
    RCGAN&{4.7775}&{4.4882}&{0.0011}&{1.4232}&{2.2441}&{0.5834}&{4.6092}&{4.4148}&{0.0276}&{1.3699}&{2.2074}&{0.6129}\\
    DDPM&{4.8053}&{4.7687}&{0.0010}&{1.5022}&{2.3843}&{0.5812}&{4.7482}&{4.4337}&{0.0279}&{1.4023}&{2.2169}&{0.6221}\\
    LDM&{4.4361}&{3.9992}&{0.0008}&{1.4083}&{1.9996}&{0.6117}&{4.1221}&{3.8972}&{0.0264}&{1.2866}&{1.9436}&{0.7432}\\
    CSDI&\underline{3.3700}&\underline{2.7227}&\underline{0.0002}&\underline{1.0086}&\underline{1.3614}&\underline{0.6676}&\underline{3.3476}&\underline{3.0105}&\underline{0.0187}&\underline{1.0017}&\underline{1.5053}&\underline{0.7990}\\
    {AppGen} &\textbf{2.6354}&\textbf{1.9678}&\textbf{0.0001}&\textbf{0.6447}&\textbf{0.9844}&\textbf{0.7893}&\textbf{2.2584}&\textbf{1.7329}&\textbf{0.0080}&\textbf{0.4476}&\textbf{0.8664}&\textbf{0.9895}\\
    \midrule
    Improvement&{21.80\%}&{27.69\%}&{50.00\%}&{36.08\%}&{27.69\%}&{18.23\%}&{32.54\%}&{42.44\%}&{57.22\%}&{55.31\%}&{42.44\%}&{23.84\%}\\
    \bottomrule
    \end{tabular}
    \label{Tab1:Shanghai}
}%
\end{table*}

\begin{table*}[tb]
\centering
\caption{App usage generation results compared with baselines on the Nanchang dataset.}
\vspace{-2mm}
\footnotesize
\resizebox{1.0\linewidth}{!}
{
    \begin{tabular}{c|ccccccccccccc}
    \toprule
    \multirow{2}{*}{\textbf{Methods}} & \multicolumn{6}{c}{{\textbf{App Popularity Distribution}}} & \multicolumn{6}{c}{\textbf{App Cate. Popularity Distribution}} \\
    \cmidrule(r){2-7}
    \cmidrule(r){8-13}
     & {RMSE} & {MAE} & {JSD} & {CRPS} & {M-TV} &{Spearmanr} & {RMSE} & {MAE} & {JSD} & {CRPS} & {M-TV} &{Spearmanr} \\
    \midrule
    Time-GAN&{5.4416}&{5.3109}&{0.0013}&{1.8683}&{0.5908}&{0.5135}&{5.1228}&{5.0023}&{0.0316}&{1.8894}&{2.5629}&{0.5159}\\
    RCGAN&{4.9686}&{4.6767}&{0.0011}&{1.4801}&{2.3393}&{0.6076}&{4.7963}&{4.5941}&{0.0287}&{1.4742}&{2.2975}&{0.5847}\\
    DDPM&{5.6703}&{5.6217}&{0.0012}&{1.7226}&{2.8135}&{0.6868}&{5.6029}&{5.2318}&{0.0329}&{1.6547}&{2.6157}&{0.6047}\\
    LDM&{4.8885}&{4.5119}&{0.0009}&{1.5913}&{2.2586}&{0.6914}&{4.6508}&{4.4338}&{0.0299}&{1.4589}&{2.1963}&{0.6972}\\
    CSDI&\underline{3.8775}&\underline{3.1312}&\underline{0.0004}&\underline{1.1597}&\underline{1.5656}&\underline{0.7677}&\underline{3.8496}&\underline{3.4813}&\underline{0.0218}&\underline{1.1514}&\underline{1.7312}&\underline{0.7578}\\
    {AppGen} &\textbf{2.8998}&\textbf{2.1606}&\textbf{0.0003}&\textbf{0.7090}&\textbf{1.0882}&\textbf{0.8672}&\textbf{2.4842}&\textbf{1.9066}&\textbf{0.0089}&\textbf{0.4932}&\textbf{0.9531}&\textbf{0.9085}\\
    \midrule
    Improvement&{25.21\%}&{31.00\%}&{25.00\%}&{38.86\%}&{30.49\%}&{12.96\%}&{35.47\%}&{45.23\%}&{59.17\%}&{57.17\%}&{44.95\%}&{19.88\%}\\
    \bottomrule
    \end{tabular}
    \label{Tab1:Nanchang}
}%
\end{table*}

\subsection{Baselines}
To evaluate AppGen, we consider the following baselines that represent the state-of-the-art technique in generation tasks, which are outlined below:

\textbf{Time-series GAN (TimeGAN)~\cite{yoon2019time}.} TimeGAN utilizes the original data as supervision and adds step-wise supervised loss, which is optimized with the adversarial objective jointly to learn the data's temporal dynamics. 

\textbf{Recurrent Conditional GAN (RCGAN)~\cite{esteban2017real}.} RCGAN leverages the recurrent neural network in the generator and the discriminator, and the RNNs are conditioned on extra auxiliary information to produce time series data. 

\textbf{Denoising Diffusion Probabilistic Models (DDPM)~\cite{ho2020denoising}.}
DDPM is based on diffusion models and is trained using a weighted variational bound designed to connect diffusion probabilistic models and denoising score matching with Langevin dynamics.
 
\textbf{Latent Diffusion Models (LDMs)~\cite{rombach2022high}.} This model applies diffusion models in the latent space of pre-trained autoencoders and introduces cross-attention layers to the model, turning the model into a conditional generator.

\textbf{Conditional Score-based Diffusion models for Imputation (CSDI)~\cite{tashiro2021csdi}.} This model employs the score-based diffusion models conditioned on the observed data via a self-supervised training method where the noise is mapped to valid information for learning interpolation targets to do the time series imputation.

%% file: Evaluations.tex
\section{Evaluation Results}\label{sec:evaluate}

\begin{figure*}[tb]
\vspace{-2mm}
\centering
\captionsetup{justification=centering}
    \begin{minipage}[t]{0.33\linewidth}
        \centering
        \includegraphics[width=1\linewidth]{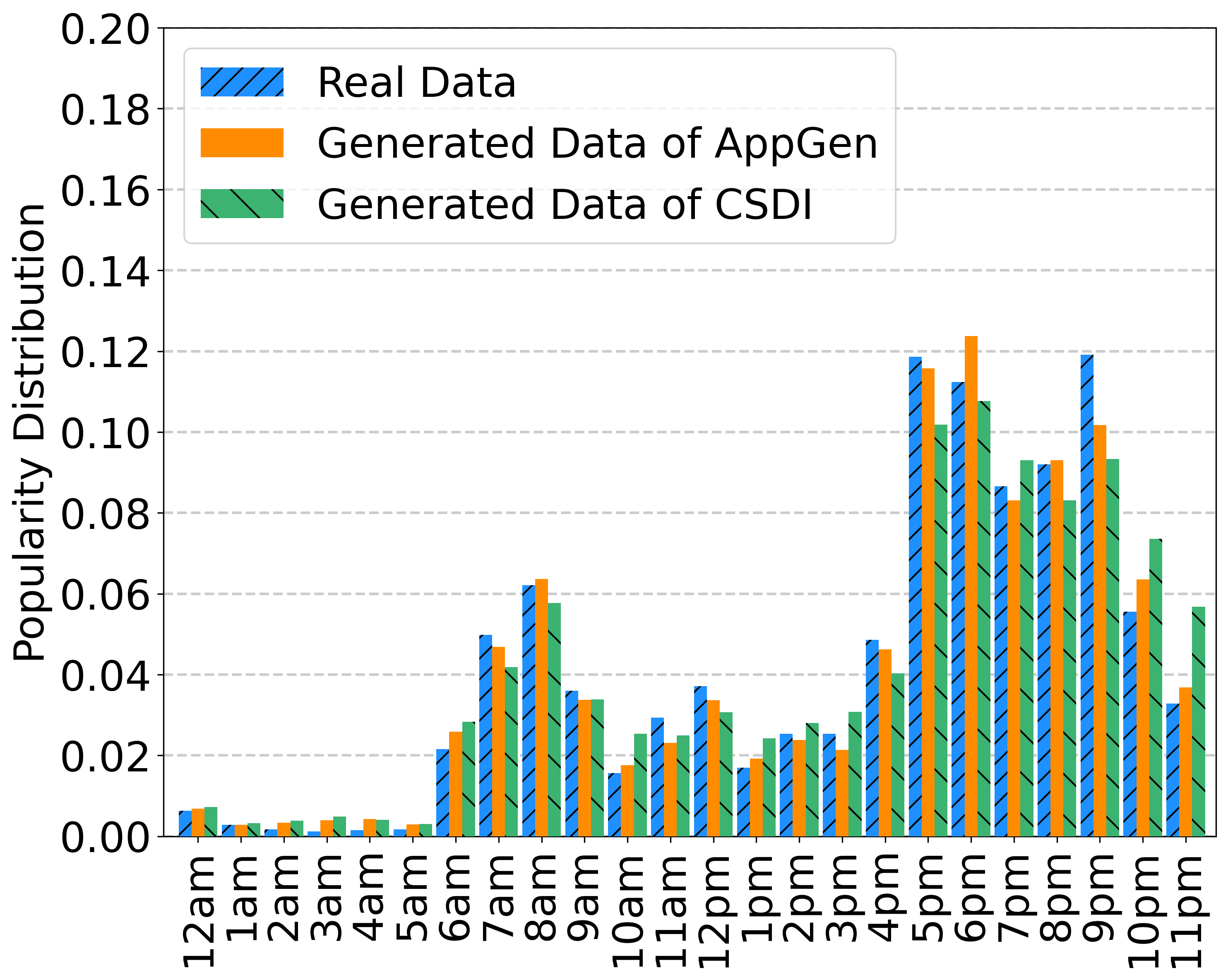}\\
        \subcaption{Music}
        \label{fig:A2-1}
    \end{minipage}%
        \hfill
    \begin{minipage}[t]{0.33\linewidth}
        \centering
        \includegraphics[width=1\linewidth]{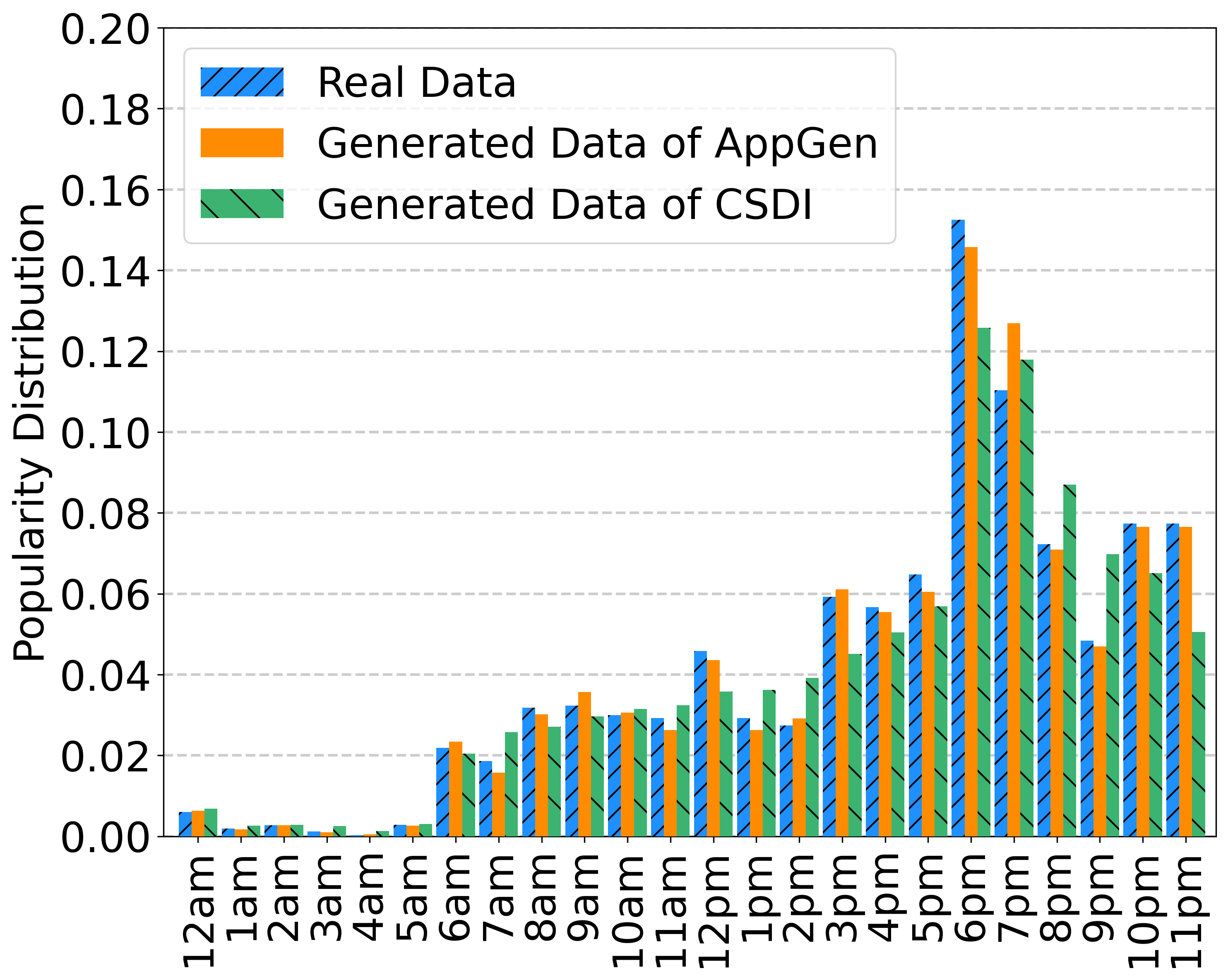}\\
        \subcaption{Games}
        \label{fig:A2-2}
    \end{minipage}%
    \hfill
    \begin{minipage}[t]{0.33\linewidth}
        \centering
        \includegraphics[width=1\linewidth]{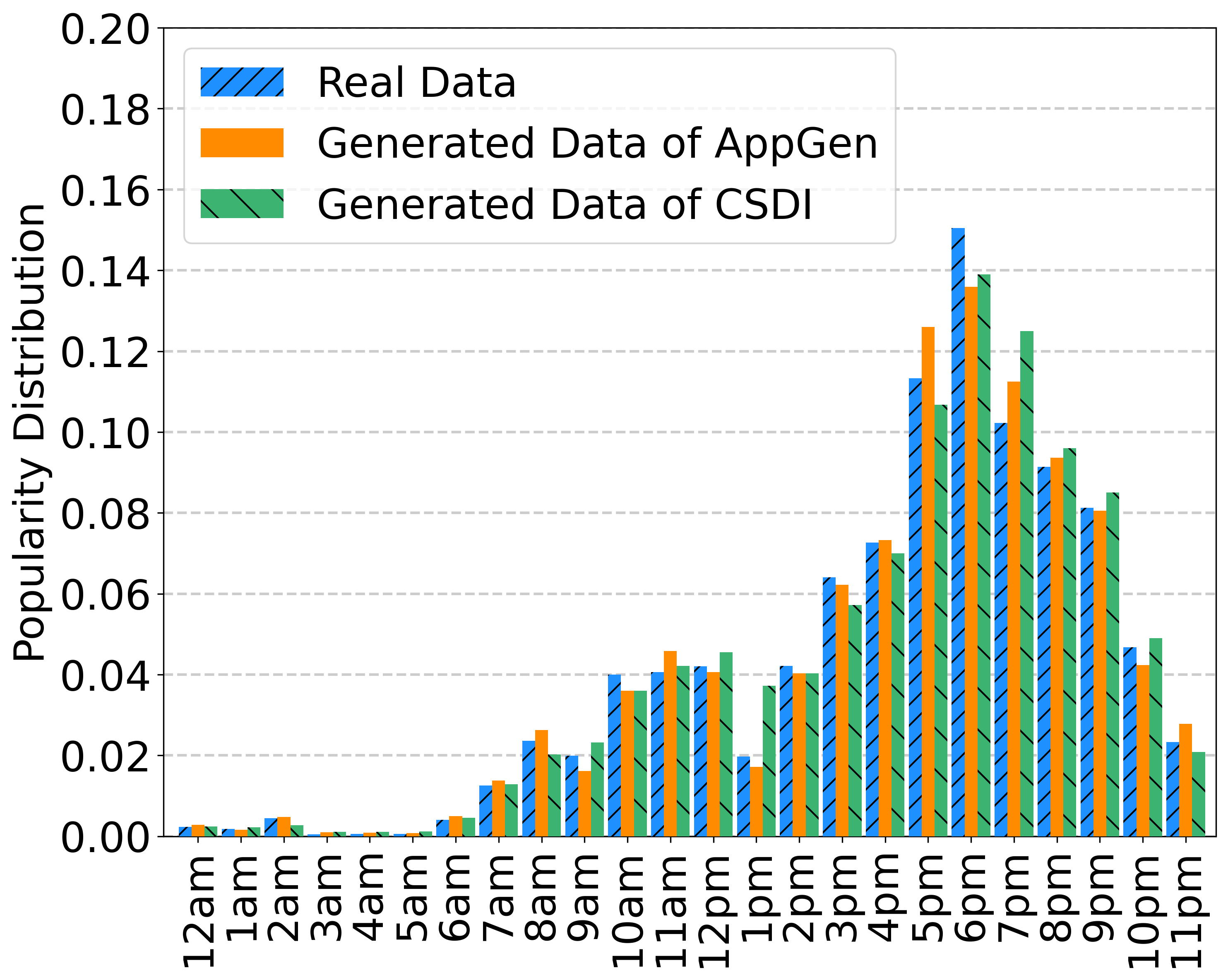}\\
        \subcaption{Lifestyle}
        \label{fig:A2-3}
    \end{minipage}%

    \medskip
    \begin{minipage}[t]{0.33\linewidth}
        \centering
        \includegraphics[width=1\linewidth]{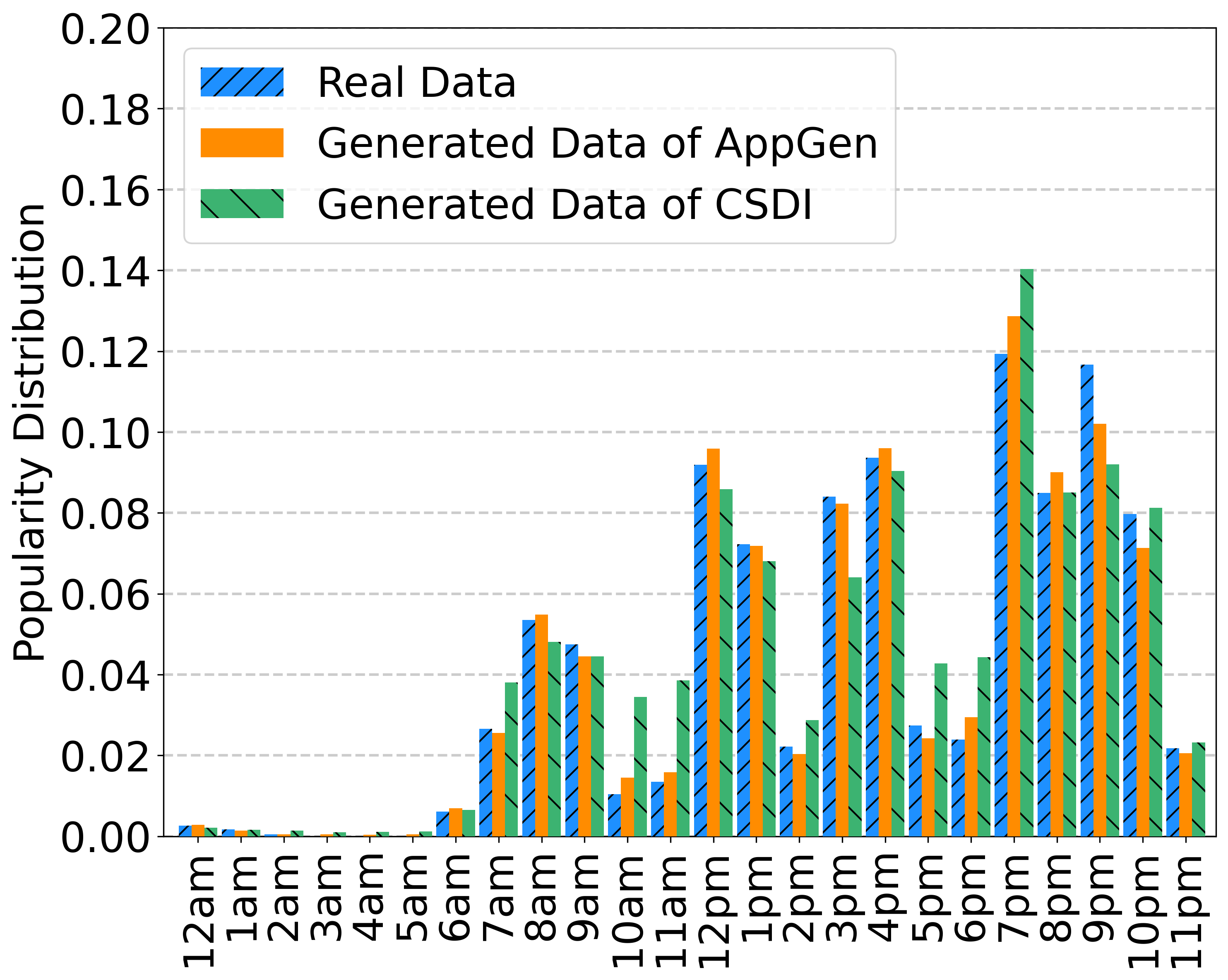}\\
        \subcaption{Books}
        \label{fig:A2-4}
    \end{minipage}%
        \hfill
    \begin{minipage}[t]{0.33\linewidth}
        \centering
        \includegraphics[width=1\linewidth]{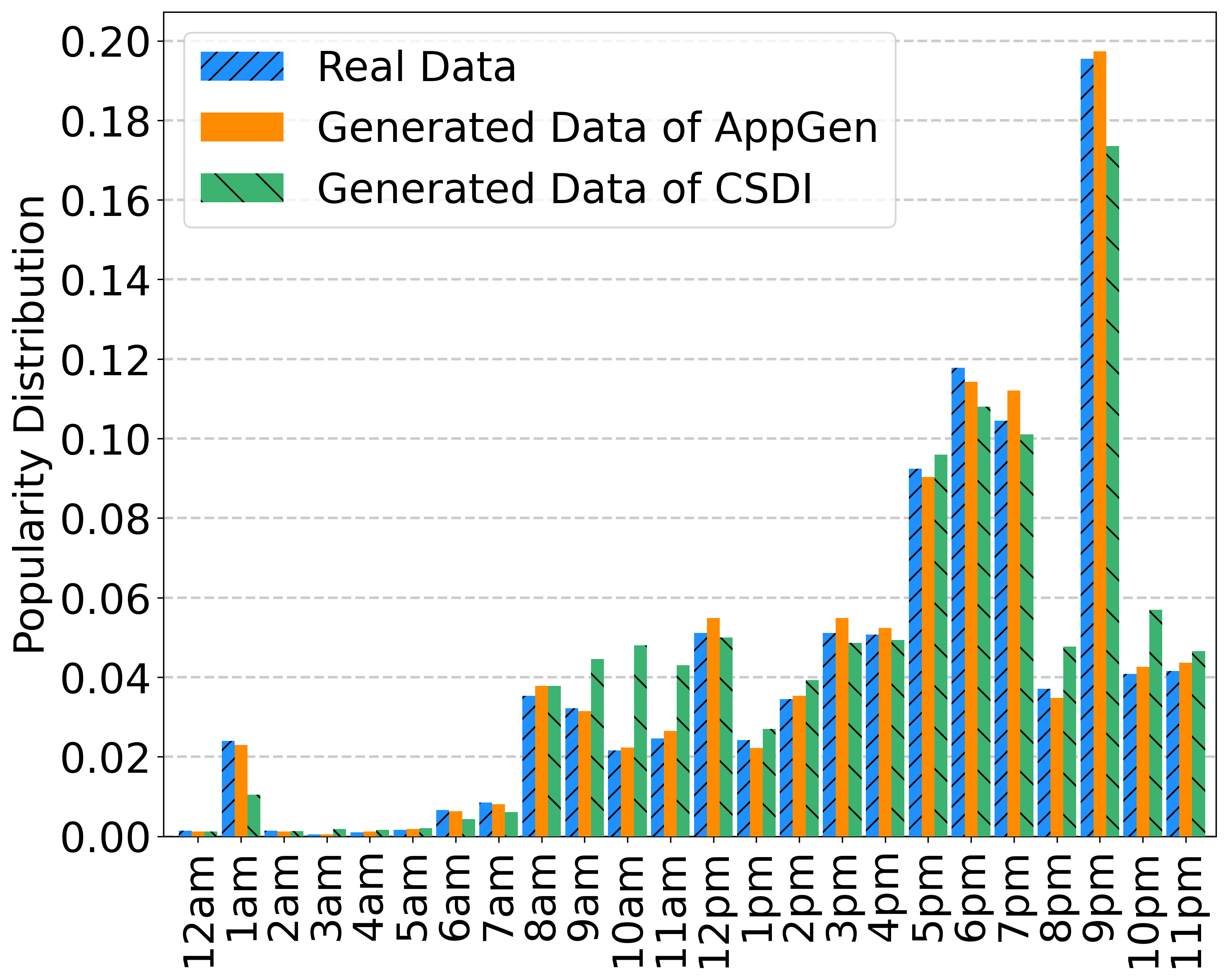}\\
        \subcaption{Shopping}
        \label{fig:A2-5}
    \end{minipage}%
    \hfill
    \begin{minipage}[t]{0.33\linewidth}
        \centering
        \includegraphics[width=1\linewidth]{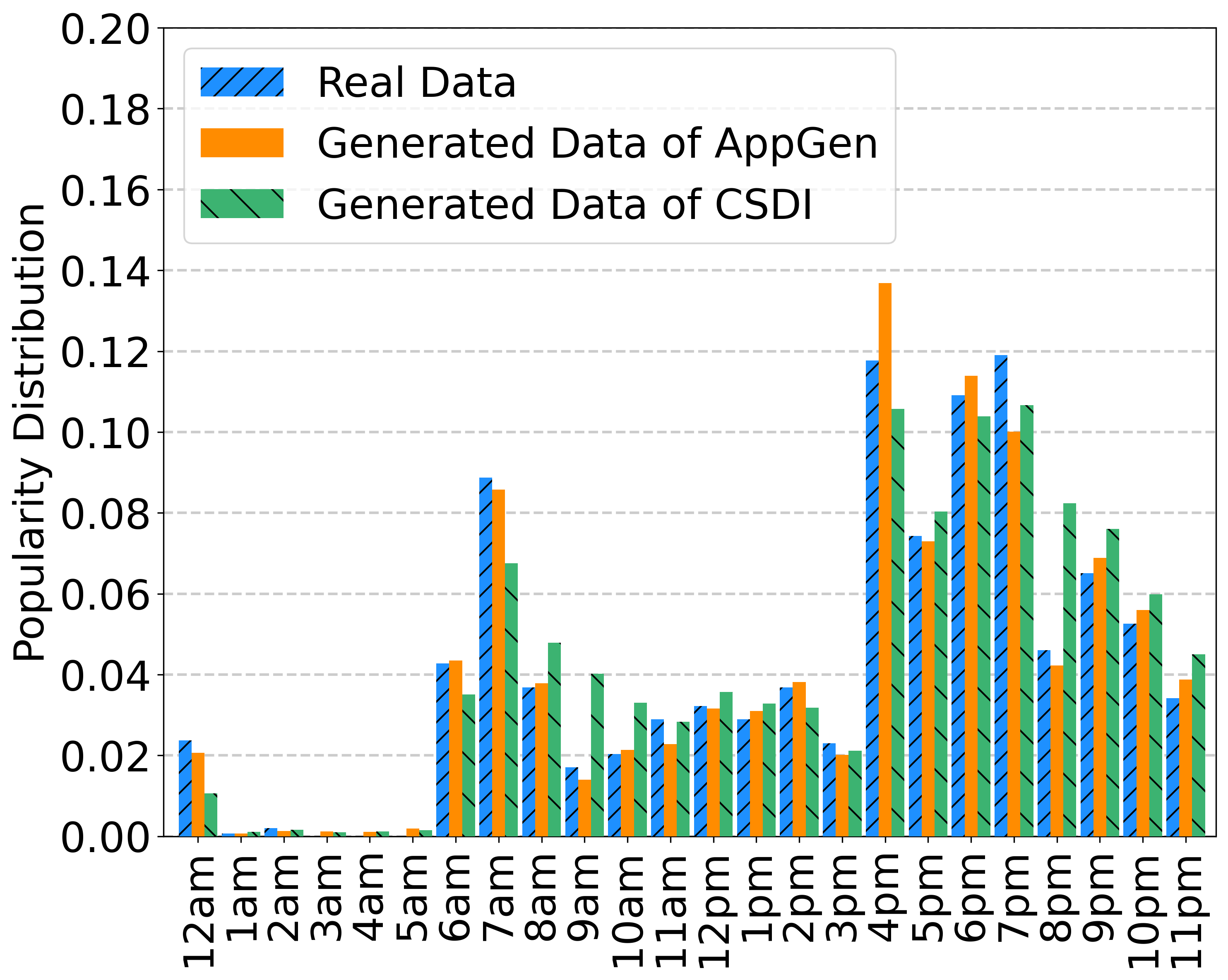}\\
        \subcaption{Weather}
        \label{fig:A2-6}
    \end{minipage}%
\vspace{-2mm}
\caption{Using popularity distribution of app category in real and generated data.}
\vspace{-2mm} 
\label{fig:time}
\end{figure*}

\begin{figure*}[tb]
\vspace{-2mm}
\centering
\captionsetup{justification=centering}
    \begin{minipage}[t]{0.33\linewidth}
        \centering
        \includegraphics[width=1\linewidth]{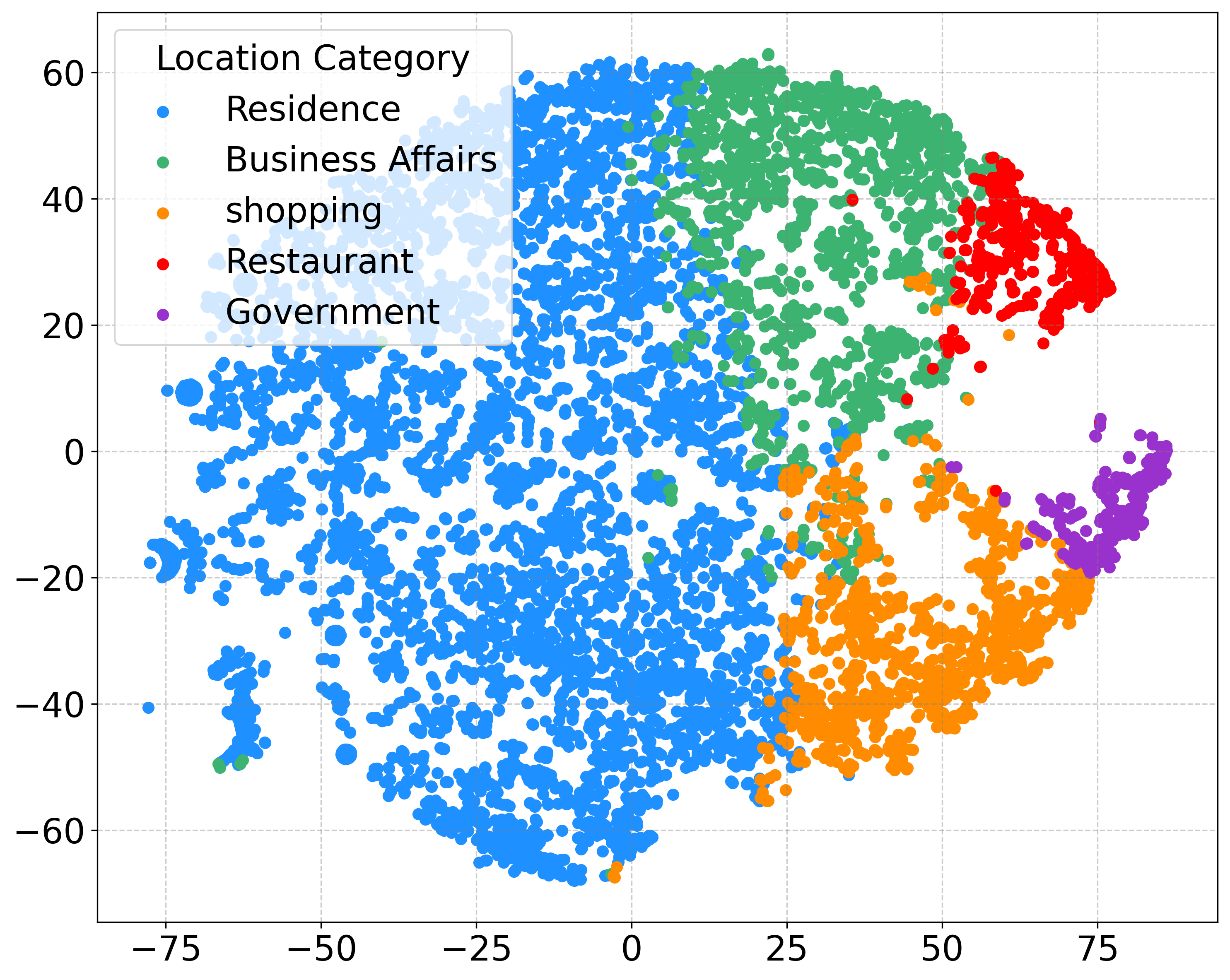}\\
        \subcaption{Real data}
        \label{fig:A3-1}
    \end{minipage}%
    \begin{minipage}[t]{0.33\linewidth}
        \centering
        \includegraphics[width=1\linewidth]{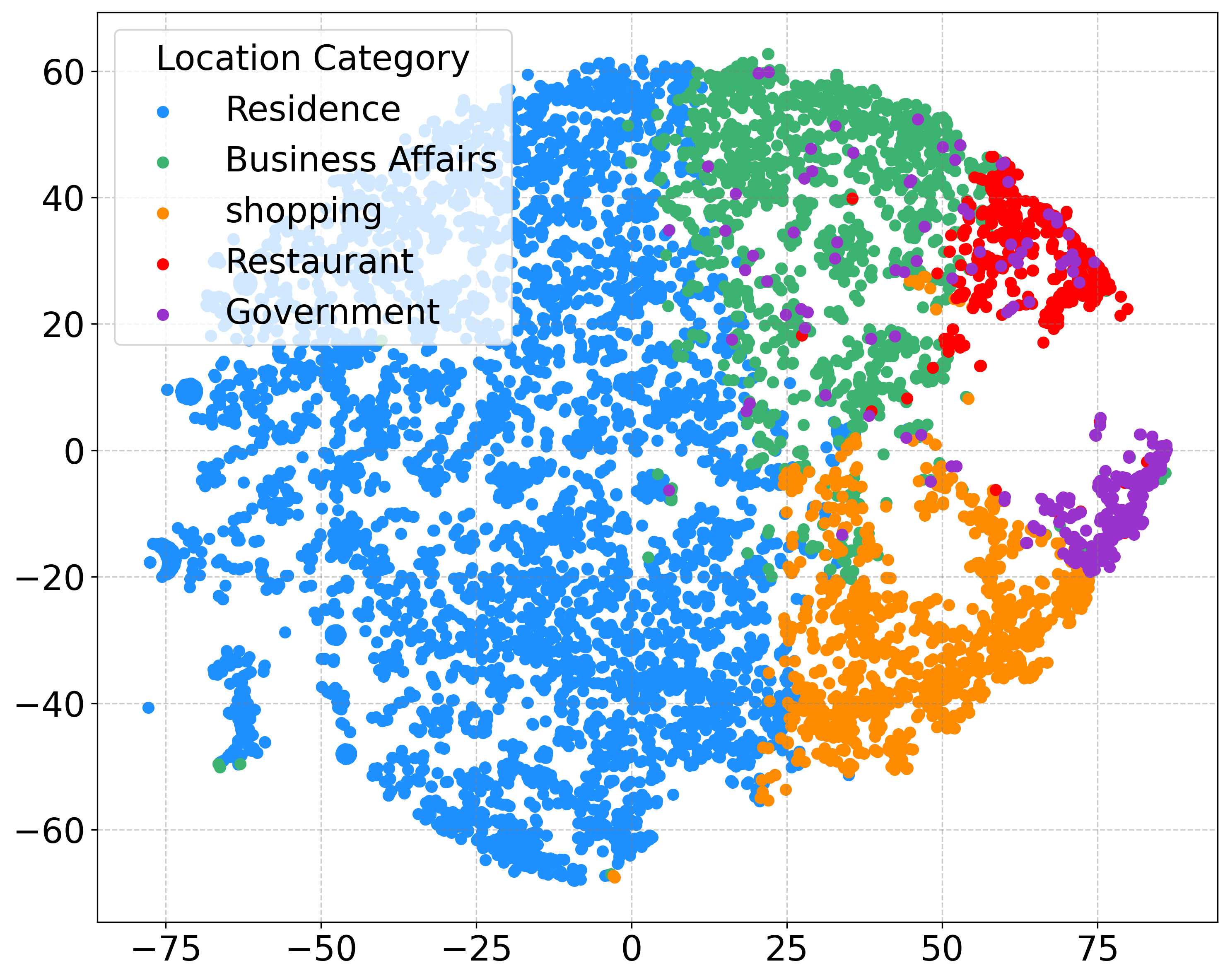}\\
        \subcaption{Generated data of AppGen}
        \label{fig:A3-2}
    \end{minipage}%
    \begin{minipage}[t]{0.33\linewidth}
        \centering
        \includegraphics[width=1\linewidth]{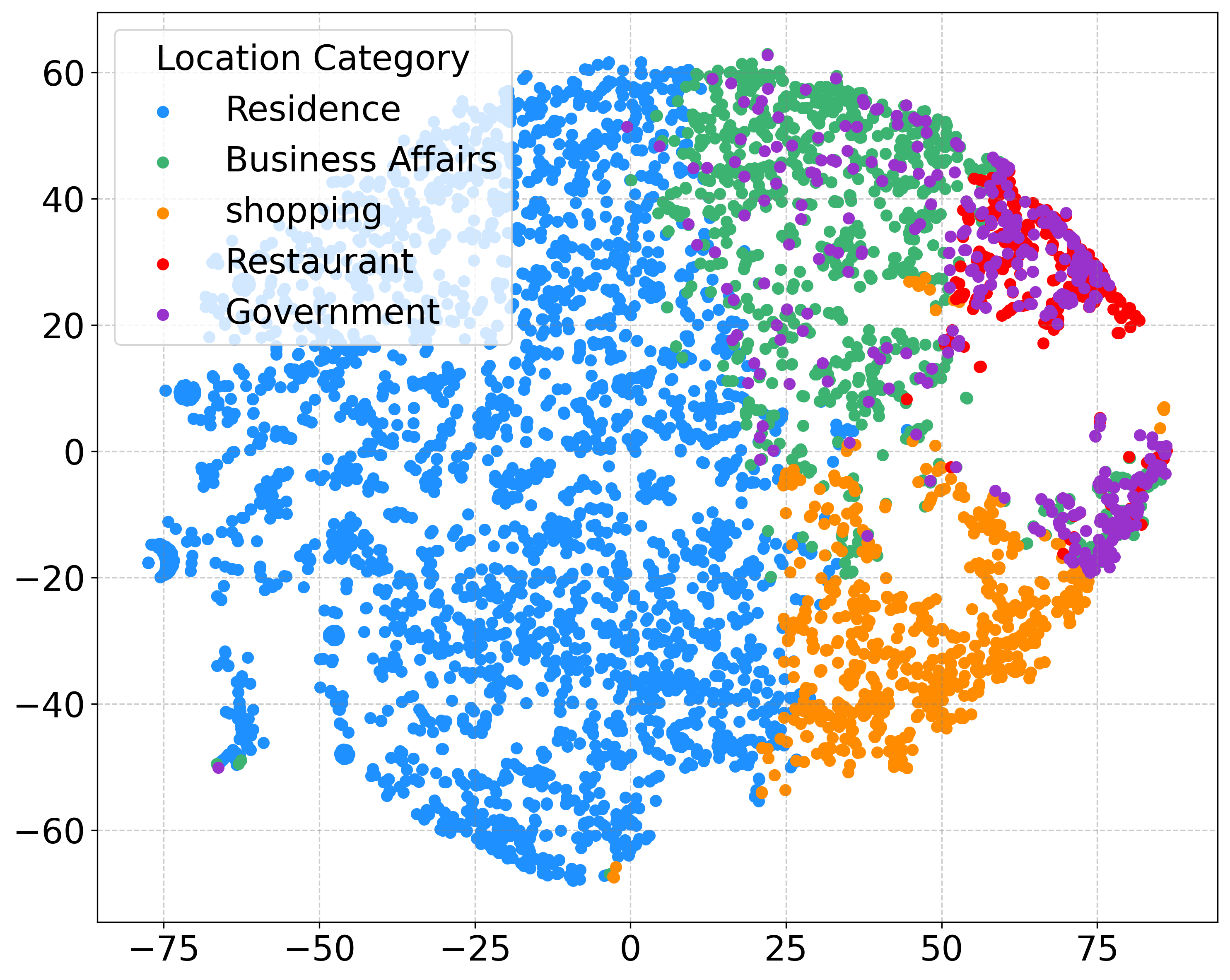}\\
        \subcaption{Generated data of CSDI}
        \label{fig:A3-3}
    \end{minipage}%
\vspace{-2mm}
\caption{Location clustering distribution in real and generated data.}
\vspace{-2mm} 
\label{fig:BS cluster}
\end{figure*}

\subsection{Overall Generation Quality}

We compare AppGen with SOTA baselines in two datasets and report the performance improvement of our model relative to the best baseline. 
As our goal is to generate the app usage sequence $\hat{A}_u$ such that its usage probability distribution matches that of the real-world usage sequence $A_u$, we measure our model from two aspects: the popularity distribution of app usage and that of app category usage. More precisely, we calculate the proportion of average daily occurrences in the app usage sequence of individual users in different apps and app categories in the real-world data and generated data as their usage popularity to be compared.

The experimental results are shown in Table~\ref{Tab1:Shanghai} and Table~\ref{Tab1:Nanchang}, which demonstrates that AppGen outperforms by all baselines over 12\% under each metric in both datasets, indicating that AppGen is superior in simulating the distribution of user's app usage. In addition, we have the following detailed observations: 

\textbf{Adding auxiliary information is necessary.} Time-GAN exhibits the lowest performance because it merely captures the temporal dynamics of data, which is insufficient for a generation. 
In contrast, RCGAN yields better results than Time-GAN, as the model is conditioned on extra auxiliary information, helping to produce time series data.

\textbf{Diffusion models can better model the uncertainty feature of behavior.}
While the two mentioned models above are based on generative adversarial network (GAN), DDPM generates the app sequence based on diffusion models under unconditional circumstances, and it demonstrates an improvement compared with Time-GAN, showing that diffusion models are more suitable in app usage generation. 

\textbf{Autoregressive method can better capture the relationship among apps.}
Compared to DDPM, LDM augments their UNet backbone of diffusion models with a conditioning mechanism, leading to improved performance. However, its architecture is mainly used for image generation, which cannot extensively capture the correlations between apps.
Compared to LDM, CSDI adopts transformer modules in the backbone of diffusion models to model the temporal dependencies of app sequence, thereby achieving a relatively competitive result. 
However, even with these enhancements, CSDI still falls short of the performance achieved by AppGen, suggesting that applying the autoregressive approach can capture the relationship among app sequences more accurately.

\subsection{Generation Quality across Temporal, Spatial and App Domains}
In addition to comparisons with various baselines, we also undertake visual comparisons between the generated sequence data and the original data.

\textbf{Generation Quality in terms of Time.}
To assess AppGen's performance in capturing the temporal features of the app usage behavior, we compare the hourly time distribution of usage intensity for various app categories across real and generated data.
\figurename~\ref{fig:time} compares time distribution across six app categories. We can find that the time distribution of the generated data is close to that of the real-world data; both have patterns of heavy app usage during the day and light usage in the early morning. Besides, despite variations in the time distribution of different app categories, generated data distribution remains relatively similar to that of original data for each app category. This proves that our model can effectively capture the general regularity of app usage behavior and specific temporal features for different app categories.

\textbf{Generation Quality in terms of Location.}
To evaluate the effectiveness of AppGen in modeling the spatial characteristics of the app usage sequence, we carry out a cluster analysis on the locations (i.e., base stations).
Specifically, we arrange the apps used in each location over time into a time series and regard the time series as a representation of each location to conduct the clustering via the $K$-$means$ algorithm. Lastly, we utilize the $t$-$SNE$ technique to reduce the dimension of location representation and visualize the clustering results.
Five location clusters are presented in \figurename~\ref{fig:BS cluster}. We can observe that the clustering results of generated data closely resemble that of real-world data, and the similarity rises with the number of locations included in the category. Hence, the proposed model can adequately model the spatial characteristics of the app usage sequence.

\begin{table*}[tb]
\centering
\caption{Frequent itemsets in real data, generated data of CSDI and generated data of AppGen.}
\vspace{-2mm}
\footnotesize
{
    \begin{tabular}{cc|cc|cc}
    \toprule
    \multicolumn{2}{c|}{{\textbf{Real Data}}} & \multicolumn{2}{c}{\textbf{Generated Data of CSDI}} & \multicolumn{2}{|c}{\textbf{Generated Data of AppGen}} \\
    \cmidrule(r){1-2}
    \cmidrule(r){3-4}
    \cmidrule(r){5-6}
    {Association Rules} & {Support} & {Association Rules} & {Support} & {Association Rules} & {Support} \\
    \midrule
    \textbf{2 $\to$ 1} & {0.032} & {148 $\to$ 1} & {0.045} & \textbf{2 $\to$ 1} & {0.049} \\
    {202 $\to$ 2} & {0.029} & {115 $\to$ 1} & {0.032} & \textbf{5 $\to$ 1} & {0.020} \\
    {4 $\to$ 1} & {0.026} & {235 $\to$ 1} & {0.022} & {17 $\to$ 1} & {0.015} \\
    \textbf{5 $\to$ 1} & {0.025} & \textbf{5 $\to$ 2} & {0.019} &  \textbf{77 $\to$ 2} & {0.015}\\
    \textbf{77 $\to$ 2} & {0.025} & \textbf{5 $\to$ 1} & {0.016} & \textbf{9 $\to$ 1} & {0.014} \\
    \textbf{9 $\to$ 1} & {0.017} & {77 $\to$ 252} & {0.016} & {77 $\to$ 1} & {0.014} \\
    {10 $\to$ 1} & {0.016} & {17 $\to$ 1} & {0.015} &  \textbf{5 $\to$ 2} & {0.013}\\
    \textbf{5 $\to$ 2} & {0.016} & {9 $\to$ 2} & {0.015} & {17 $\to$ 2} & {0.011}\\
    {202 $\to$ 7} & {0.015} & {115 $\to$ 2} & {0.015} & {9 $\to$ 2} & {0.010}\\
    \textbf{22, 7 $\to$ 1} & {0.015} & {148 $\to$ 2} & {0.012} & \textbf{22, 7 $\to$ 1} & {0.010} \\
    \bottomrule
    \end{tabular}
}%
\label{Tab3:itemset}
\vspace{-2mm}
\end{table*}

\begin{table*}[tb]
\centering
\caption{Evaluation results generated by different variants of AppGen on the Shanghai dataset.}
\vspace{-2mm}
\footnotesize
\resizebox{1.0\linewidth}{!}
{
    \begin{tabular}{c|ccccccccccccc}
    \toprule
    \multirow{2}{*}{\textbf{Methods}} & \multicolumn{6}{c}{{\textbf{App Popularity Distribution}}} & \multicolumn{6}{c}{\textbf{App Cate. Popularity Distribution}} \\
    \cmidrule(r){2-7}
    \cmidrule(r){8-13}
     & {RMSE} & {MAE} & {JSD} & {CRPS} & {M-TV} &{Spearmanr} & {RMSE} & {MAE} & {JSD} & {CRPS} & {M-TV} &{Spearmanr} \\
    \midrule
    {w/o spatial context}&{2.9712}&{2.1024}&{0.0001}&{0.6993}&{1.0056}&{0.7488}&{2.6767}&{1.9011}&{0.0163}&{0.5167}&{0.9789}&{0.9013}\\
    {w/o historical behavioral data}&{4.3526}&{3.9648}&{0.0003}&{0.9965}&{1.9824}&{0.6521}&{3.3089}&{3.2122}&{0.0372}&{0.8212}&{1.6061}&{0.8364}\\
    {w/o current spatio-temporal context}&{3.3546}&{2.3401}&{0.0003}&{0.8340}&{1.1701}&{0.6690}&{3.1562}&{2.3820}&{0.0272}&{0.7382}&{1.1910}&{0.8571}\\
    {AppGen} &\textbf{2.6354}&\textbf{1.9678}&\textbf{0.0001}&\textbf{0.6447}&\textbf{0.9844}&\textbf{0.7893}&\textbf{2.2584}&\textbf{1.7329}&\textbf{0.0080}&\textbf{0.4476}&\textbf{0.8664}&\textbf{0.9895}\\
    \bottomrule
    \end{tabular}
    \label{Tab2:component}
}%
\vspace{-2mm}
\end{table*}

\textbf{Frequent Itemsets Statistics of Apps.}
To measure the performance of AppGen in capturing the temporal relationship between apps in the sequence, we utilize the $Apriori$ algorithm to count the frequent itemsets of apps in real-world data, generated data of CSDI, and generated data of AppGen, respectively. The result is demonstrated in Table~\ref{Tab3:itemset}. We can observe that the frequent itemsets of the data generated by AppGen match better with real-world data, and the ranking of matching items with support is consistent. This proves that by applying the autoregressive method, our model can well learn and model the temporal relationship between apps.

\subsection{Ablation Study}
To demonstrate the effectiveness of the designs we adopted, we evaluate the performance of AppGen with the following three variants:

\textbf{w/o spatial context.} In this variant, we remove the spatial contextual information and merely conduct the app usage generation conditionally on having temporal context. 

\textbf{w/o historical behavioral data.} This variant removes the historical behavioral features and simply uses the current spatio-temporal context as the condition to guide app usage generation.

\textbf{w/o current spatio-temporal context.} In this variant, we solely regard the past behavioral representation as a condition for app usage generation, eliminating the current spatio-temporal information.

The experimental results are presented in Table~\ref{Tab2:component}. 
We can observe that the model's performance unconditioned with historical behavioral information is the worst in all metrics. This suggests that it is imperative to consider the user's historical app usage behavior when conducting generation. 
Additionally, removing the current spatio-temporal context makes the model underperform AppGen, indicating that it is essential to consider the user's current visited location and time.
At last, without spatial contextual information, the model fails to capture the correlation between app usage behavior and spatial information, thus resulting in a relatively poor performance in app usage generation.

%% file: Application.tex
\section{Application Use Cases}
\label{sec:application}
We further conduct the app usage behavior prediction to verify the effectiveness of our model. Specifically, we conduct experiments on five classic prediction algorithms using three different dataset combinations, and we want to confirm that our generated data can be generalized to practical case applications.

\begin{figure}[tb]
\vspace{-2mm}
\centering
\includegraphics[width=0.7\linewidth]{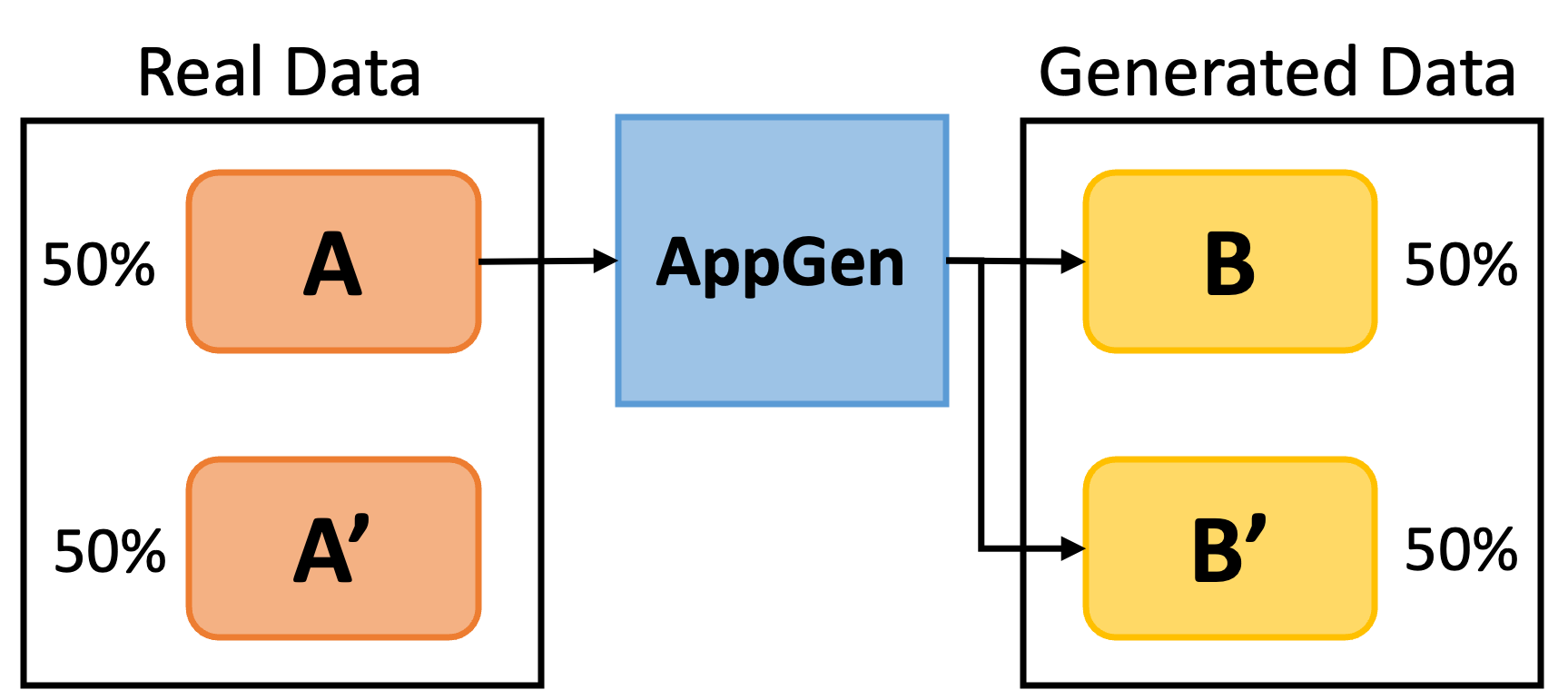}
\caption{App usage prediction modelling data setup.}
\label{fig:datasplit}
\vspace{-2mm}
\end{figure}

We first partition our dataset, as shown in \figurename~\ref{fig:datasplit}. 
The real-world data is split into two equal parts: a training set $A$ and a test set $A'$. We then train AppGen using training set $A$ and generate datasets $B$ and $B'$ for further training and testing. Finally, we evaluate different app usage prediction algorithms by training on $A$ and/or $B$ and testing on $A'$ or $B'$, to examine the generalization capability of the generated data. The findings are summarized in Table~\ref{Tab4:prediction}.

\textbf{Algorithm Comparison.} We first assess whether the rankings of app usage prediction algorithms are consistent when using generated data. In this scenario, we only utilize data generated by AppGen. Our goal is to see if the ranking of different predictive algorithms on real-world data is maintained when trained and tested on the generated data. Specifically, we train each predictive model on dataset $B$ and test on $B'$ (Exp\#2), then compare these rankings with the ground truth rankings, where models are trained on $A$ and tested on $A'$ (Exp\#1). The results show that the performance rankings of the five algorithms are consistent across both real-world and generated data, indicating that the generated data is effective in differentiating between the various algorithms.

\textbf{Data Augmenting Ability.} Besides, we assess whether generated data can have a data augmentation effect on real-world data. In this instance, we train the predictive models on $A$ and $B$, test on $B'$ (Exp\#3), and compare the results with the ground truth prediction (Exp\#1). We can find that adding generated data to the original dataset leads to a clear performance improvement between Exp\#1 and Exp\#3, indicating that the generated data has the effect of data augmentation.

\begin{table*}[tb]
\centering
\caption{App usage prediction results by different models on different datasets.}
\vspace{-2mm}
\footnotesize
\resizebox{1.0\linewidth}{!}
{
    \begin{tabular}{c|c|ccccccccccccccc}
    \toprule
    \multirow{2}{*}{} & \multirow{2}{*}{\textbf{Methods}} & \multicolumn{3}{c}{{\textbf{Acc@k}}} & \multicolumn{3}{c}{{\textbf{MRR@k}}} & \multicolumn{3}{c}{{\textbf{NDCG@k}}} & \multicolumn{3}{c}{{\textbf{Recall@k}}} & \multicolumn{3}{c}{{\textbf{F1 score@k}}}\\
    \cmidrule(r){3-5}
    \cmidrule(r){6-8}
    \cmidrule(r){9-11}
    \cmidrule(r){12-14}
    \cmidrule(r){15-17}
    & & {1} & {5} & {10} & {1} & {5} & {10} & {1} & {5} & {10} & {1} & {5} & {10} & {1} & {5} & {10}\\
    \midrule
    \multirow{5}{*}{Exp\#1} & {LSTM~\cite{hochreiter1997long}} & {0.2056} & {0.3238} & {0.3822} & {0.2056} & {0.2806} & {0.2927} & {0.2056} & {0.3156} & {0.3413} & {0.3261} & {0.3823} & {0.4583} & {0.2522} & {0.3506} & {0.4168} \\
    & {Transformer} & {0.2234} & {0.3297} & {0.3901} & {0.2234} & {0.2923} & {0.2956} & {0.2234} & {0.3364} & {0.3448} & {0.3023} & {0.4055} & {0.4790} & {0.2569} & {0.3637} & {0.4300} \\
    & {CAP~\cite{chen2019cap}} & {0.2260} & {0.3446} & {0.3994} & {0.2260} & {0.2978} & {0.3082} & {0.2260} & {0.3457} & {0.3623} & {0.3009} & {0.4379} & {0.4822} & {0.2581} & {0.3857} & {0.4369} \\
    & {SA-GCN~\cite{yu2020semantic}} & {0.2422} & {0.4470} & {0.5189} & {0.2422} & {0.3207} & {0.3305} & {0.2422} & {0.3524} & {0.3755} & {0.3322} & {0.4911} & {0.5643} & {0.2801} & {0.4680} & {0.5406} \\
    & {AppUsage2Vec~\cite{zhao2019appusage2vec}} & {0.2555} & {0.5678} & {0.6896} & {0.2555} & {0.3669} & {0.3843} & {0.2555} & {0.4177} & {0.4590} & {0.3492} & {0.5831} & {0.6907} & {0.2950} & {0.5753} & {0.6901} \\
    \midrule
    \multirow{5}{*}{Exp\#2} & {LSTM} & {0.2076} & {0.3365} & {0.3944} & {0.2076} & {0.2842} & {0.3128} & {0.2076} & {0.3379} & {0.3631} & {0.3189} & {0.3902} & {0.4832} & {0.2515} & {0.3614} & {0.4339} \\
    & {Transformer} & {0.2305} & {0.3412} & {0.3987} & {0.2305} & {0.3111} & {0.3134} & {0.2305} & {0.3529} & {0.3509} & {0.3043} & {0.4192} & {0.4843} & {0.2623} & {0.3762} & {0.4374} \\
    & {CAP} & {0.2331} & {0.3699} & {0.4072} & {0.2331} & {0.3121} & {0.3189} & {0.2331} & {0.3502} & {0.3782} & {0.3110} & {0.4991} & {0.4920} & {0.2664} & {0.4057} & {0.4456} \\
    & {SA-GCN} & {0.2465} & {0.4689} & {0.5379} & {0.2465} & {0.3390} & {0.3592} & {0.2465} & {0.3812} & {0.3920} & {0.3299} & {0.5023} & {0.5823} & {0.2822} & {0.4850} & {0.5592} \\
    & {AppUsage2Vec} & {0.2623} & {0.5903} & {0.6931} & {0.2623} & {0.3749} & {0.4024} & {0.2623} & {0.4345} & {0.4623} & {0.3611} & {0.6039} & {0.7039} & {0.3038} & {0.5970} & {0.6985} \\
    \midrule
    \multirow{5}{*}{Exp\#3} & {LSTM} & {0.2133} & {0.3279} & {0.3923} & {0.2133} & {0.2836} & {0.2984} & {0.2133} & {0.3264} & {0.3580} & {0.3178} & {0.3829} & {0.4729} & {0.2553} & {0.3533} & {0.4288} \\
    & {Transformer} & {0.2279} & {0.3303} & {0.3978} & {0.2279} & {0.3023} & {0.3092} & {0.2279} & {0.3440} & {0.3492} & {0.3115} & {0.4073} & {0.4847} & {0.2632} & {0.3648} & {0.4370} \\
    & {CAP} & {0.2343} & {0.3520} & {0.4045} & {0.2343} & {0.3024} & {0.3146} & {0.2343} & {0.3491} & {0.3789} & {0.3007} & {0.4472} & {0.4927} & {0.2633} & {0.3939} & {0.4443} \\
    & {SA-GCN} & {0.2516} & {0.4524} & {0.5280} & {0.2516} & {0.3328} & {0.3528} & {0.2516} & {0.3836} & {0.3921} & {0.3288} & {0.5023} & {0.5683} & {0.2850} & {0.4760} & {0.5474} \\
    & {AppUsage2Vec} & {0.2631} & {0.5729} & {0.6925} & {0.2631} & {0.3735} & {0.3941} & {0.2631} & {0.4322} & {0.4672} & {0.3512} & {0.5863} & {0.7042} & {0.3008} & {0.5795} & {0.6983} \\
    \bottomrule
    \end{tabular}
}%
\caption*{\raggedright 
Exp\#1: training set--$A$, testing set--$A'$;

Exp\#2: training set--$B$, testing set--$B'$;

Exp\#3: training set--$A+B$, testing set--$A'$.
} 
\label{Tab4:prediction}
\vspace{-2mm}
\end{table*}

%% file: Related.tex
\section{Related Work}
\label{sec:related}

\subsection{App Usage Prediction and Recommendation}
Unlike the task of app usage generation, the primary goal of app usage prediction is to recommend the next app that users are most likely to use based on their prior app usage data. In recent years, numerous studies have explored using contextual information, such as the user's current status, location, and time of day, to improve prediction performance. For instance, Lu~\emph{et al.}~\cite{lu2018mining} proposed a framework that adopted location and temporal segments as the context in prediction. Kaji~\emph{et al.}~\cite{kaji2011app} developed AppLocky, which requests users to select their current context for context-aware app recommendations.

In addition to considering contextual data, several researchers apply app usage sequences to study app usage prediction. Given the sequences, some studies predict app usage based on Markovian property. Natarajan~\emph{et al.}~\cite{natarajan2013app} introduced a cluster-level Markov model, which personally forecasts the usage of specific apps. Some studies also seek to capture app usage patterns from historical app usage sequences. For instance, Hwang~\emph{et al.}~\cite{hwang2019mobile} introduced the App-Usage Tracking Feature which characterizes each app from the previously used app sequence, and Katsarou~\emph{et al.}~\cite{katsarou2022whatsnextapp} proposed WhatsNextApp, an approach based on LSTM (Long Short-Term Memory) networks by using sequences of app usage logs.

Furthermore, other studies conduct app usage prediction by employing graph-based representation learning methods, seeking to model the correlations between apps via constructing graphs based on prior app usage data. Ouyang~\emph{et al.}~\cite{ouyang2022learning} proposed DUGN, adopting the dynamic graph structure and using the hierarchical graph attention mechanism to recommend the next app to be used. Fang~\emph{et al.}~\cite{fang2024enhancing} introduced MP-GT, which leverages Graph Convolutional Network (GCN) and Transformer techniques to recommend the next app that a user is most likely to use.

\subsection{Generative Models}
Generative models are machine learning models that aim to learn the underlying patterns or distributions of the input data through iterative training and then generate new synthetic data. Generative models come in various forms, each with its unique approach to understanding and generating data. Ian~\emph{et al.}~\cite{goodfellow2014generative} proposed Generative Adversarial Networks (GANs), which consist of two neural networks, the generator and the discriminator. The generator tries to produce data, while the discriminator attempts to distinguish between real and generated data. In the manner of a zero-sum game, the generator and discriminator contest with one other and are trained together until the generator improves to the point where the discriminator is unable to detect any difference. In contrast to the operation mechanism of GANs, the variational autoencoder (VAE)~\cite{kingma2013auto} generates data by sampling from a learned probability distribution. It comprises an encoder and a decoder. The encoder transforms the input data into a latent space vector, while the decoder reconstructs the data from these low-dimensional representations to produce new data. During optimization, the loss function minimizes the disparity between the original and generated data while ensuring that the latent representations resemble a standard normal distribution. Diffusion models can be viewed as a hierarchical VAE network~\cite{luo2022understanding}. They employ a Markov chain to recreate the original data distribution, generating data by progressively denoising noisy samples along the chain.

%% file: Conclusion.tex
\section{Conclusion}
\label{sec:conclude}
This paper proposes AppGen, an autoregressive diffusion model designed to synthesize personalized app usage behavior data based on mobility trajectories. The model incorporates several innovative aspects. AppGen leverages probabilistic diffusion models to capture the inherent uncertainty of app usage behavior. It applies autoregression for generation, utilizing multiple techniques within the autoregressive module to extract and capture correlations among apps in the sequence. Additionally, AppGen constructs an urban knowledge graph to obtain well-learned semantic representations of the spatio-temporal context. We evaluate AppGen using two real-world app usage behavior datasets. Our results demonstrate that AppGen significantly outperforms existing approaches in synthesizing app usage behavior data that matches the real-world data distribution and is also effective in practical application use cases.